%% file: main.tex
\documentclass[sigconf]{acmart}
\usepackage{graphicx}
\usepackage{caption}
\usepackage{subcaption}
\usepackage{tabularx}
\usepackage{array}
\usepackage[table,xcdraw]{xcolor} 
\usepackage{pdflscape} 
\AtBeginDocument{%
  }


\copyrightyear{2026}
\acmYear{2026}
\setcopyright{cc}
\setcctype{by}
\acmConference[ASSETS '26]{The 28th International ACM SIGACCESS Conference on Computers and Accessibility}{October 25--28, 2026}{Vila Nova de Gaia, Portugal}
\acmBooktitle{The 28th International ACM SIGACCESS Conference on Computers and Accessibility (ASSETS '26), October 25--28, 2026, Vila Nova de Gaia, Portugal}
\acmDOI{10.1145/3797867.3829020}
\acmISBN{979-8-4007-2521-0/2026/10}

\begin{document}

\title["I Don't Want My Mental Health App To Give Me Mental Health Barriers"]{"I Don't Want My Mental Health App To Give Me Mental (Health) Barriers": Unpacking The Need For Digital Mental Health Tracking Services With And For The Blind Community}

\author{Omar Khan}
\email{mkhan259@illinois.edu}
\orcid{0009-0005-3209-3525}
\affiliation{%
    \department{Siebel School of Computing and Data Science}
    \institution{University of Illinois Urbana-Champaign}
    \city{Urbana}
    \state{Illinois}
    \country{USA}
}

\author{JooYoung Seo}
\email{jseo1005@illinois.edu}
\orcid{0000-0002-4064-6012}
\affiliation{%
    \department{School of Information Sciences}
    \institution{University of Illinois Urbana-Champaign}
    \city{Champaign}
    \state{Illinois}
    \country{USA}
}

\input{chapter/0_abstract}




\maketitle

\input{chapter/1_introduction}
\input{chapter/2_related_work}
\input{chapter/3_methods}
\input{chapter/4_findings}
\input{chapter/5_discussion}
\input{chapter/6_limitations}

\input{chapter/7_conclusion}
\input{chapter/9_acknowledgements}

\bibliographystyle{ACM-Reference-Format}
\bibliography{references/references, references/a11y_framework, references/software}

\input{chapter/10_appendix}

\end{document}

%% file: chapter/0_abstract.tex
\begin{abstract}
\label{sec:abstract}

Digital mental health (DMH) tracking services promise continuous, personalized support for well-being, but their design often assumes sighted users. For the blind community, this assumption produces a distinct pattern of exclusion: services whose accessibility cannot be evaluated without first paying for them, community features that exclude the users they purport to support, and interfaces that leave users digitally literate but functionally blocked. We report on an explanatory sequential mixed-methods study of blind users' experiences with DMH tracking services in the United States. In the first phase, 93 legally blind adults completed a survey about their usage patterns, adoption decisions, and data-agency preferences; in the second, 10 survey respondents participated in semi-structured interviews. We analyzed closed-ended responses using descriptive statistics and the Kruskal–Wallis test, and open-ended and interview data using inductive thematic analysis, interpreting findings through Norman and Skinner's eHealth Literacy framework. Participants identified mindfulness, sleep, and goal-tracking services as their most-used categories, but also described recurring exclusion from the community-support features that other users value most. We argue that the framework's "computer literacy" dimension is insufficient on its own: many of our participants possessed the literacy but were blocked from applying it by design choices that predate the user. We contribute design recommendations for transparent pre-purchase accessibility evaluation, accessibility-native rather than retrofitted interfaces, and user-controlled data agency —- recommendations intended not to accommodate blind users but to design DMH tracking services with them from the start.

\end{abstract}

%% file: chapter/1_introduction.tex
\section{Introduction}
\label{sec:introduction}

Digital mental health (DMH) tracking services now offer a wide range of supports, including stress and anxiety management, mood tracking, mindfulness and meditation, medication and symptom management, and cognitive exercise~\cite{ayobiDigitalMentalHealth2022, baghaeiTimeGetPersonal2020, oewelApproachesTailoringBetweensession2024}. Prior work has documented their benefits for helping users identify thought patterns, form habits, and track change over time~\cite{bowie-dabreoUserPerspectivesEthical2022}. Yet, the design of these services can make assumptions about a user's abilities: reading visual-centric charts, tapping precisely located buttons, and interpreting color-coded feedback. For the blind community, this assumption translates into exclusion at multiple points: screen-reader incompatibility~\cite{soubutts_challenges_2024, choi_exploring_2024, leeIdentifyAdaptPersist2024, lee_personal_2023}, inaccessible data visualizations~\cite{lee_personal_2023}, and in some cases, inaccessibility that cannot be discovered until after payment.
 
The stakes for blind users are especially high; blind individuals report higher rates of depression and anxiety than the general population~\cite{kohdaMentalHealthStatus2019, richardsonUnderutilizationMentalHealth2024} and face systemic barriers to in-person mental health care, including limited provider cultural competency and insufficient institutional accommodation~\cite{mcdonnall_availability_2017}. DMH tracking services could, in principle, expand access for a community that is underserved by in-person services. Instead, they often reproduce the same exclusions in digital form, while introducing new ones specific to the DMH
marketplace.
 
Recent HCI work has begun to examine blind users' engagement with personal health technologies more broadly~\cite{choi_exploring_2024, leeIdentifyAdaptPersist2024, lee_personal_2023, millerSelfMonitoringPhysicalActivity2022}, and Khan and Seo~\cite{khan_sighted_2025} partnered with blind advocacy organizations to surface initial DMH-specific needs. Outside of this work, systematic investigation of DMH accessibility with and for the blind community remains limited. In particular, we lack evidence on: which DMH categories blind users actually engage with and why; how they make adoption decisions when accessibility cannot be evaluated before purchase; and what they want from the data their DMH services collect.
 
We investigated these questions through an explanatory sequential mixed-methods design~\cite{creswell_designing_2017}, combining a cross-sectional survey ($n=93$) with follow-up semi-structured interviews ($n=10$). Three research questions (RQs) guided our work:
 
\begin{itemize}
  \item \textbf{RQ1:} What needs exist for DMH tracking services among the blind community?
  \item \textbf{RQ2:} What factors influence the adoption of DMH tracking services among blind users?
  \item \textbf{RQ3:} How can data management features in DMH tracking services better support data agency for blind users?
\end{itemize}

We interpret our findings through Norman and Skinner's eHealth Literacy framework~\cite{norman_ehealth_2006, norgaardEhealthLiteracyFramework2015}, which identifies six literacy domains: traditional, health, information, scientific, media, and computer, required for effective eHealth engagement. The framework has been widely applied in digital health scholarship, but we argue it is incomplete for the accessibility context: our participants frequently possessed the literacies in question but were structurally blocked from exercising them. We therefore extend the framework by distinguishing between \textit{literacy gaps} (where users lack a competency) and \textit{access barriers} (where design choices render existing competencies unusable).
 
Our contributions are four-fold:
 
\begin{itemize}
  \item \textbf{Empirical findings on blind users' DMH tracking service use} at a large-scale, including category-level usage, ratings of accessibility and helpfulness, and reasons for non-adoption.
  \item \textbf{A characterization of paywall-gated accessibility evaluation}, a structural feature of the DMH marketplace that forces blind users to commit financial resources before they can assess whether a service is usable.
  \item \textbf{Evidence that inaccessible DMH services compound MH-specific isolation} by excluding blind users from the peer-support and community features that draw other users in. 
  \item \textbf{A conceptual extension of the eHealth Literacy framework} that distinguishes literacy gaps from access barriers, with implications for how accessibility research reads and applies health literacy models.
\end{itemize}

While building upon the preliminary survey results detailed in Khan and Seo \cite{khan_sighted_2025}, this paper integrates a new follow-up interview phase ($n = 10$), addresses research questions through a unified analysis of both phases, and proposes a conceptual extension to the eHealth Literacy framework~\cite{norman_ehealth_2006, norgaardEhealthLiteracyFramework2015}; the original survey data serves primarily as context for these new analyses.

The remainder of the paper is structured as follows. Section~\ref{sec:related_work} situates our work in HCI scholarship on mental health technology and accessibility. Section~\ref{sec:methods} details our mixed-methods design, including how we mapped our instruments to TAM and CSQ constructs. Section~\ref{sec:findings} presents findings organized by research question, integrating survey and interview data within each. Section~\ref{sec:discussion} discusses implications for DMH design, policy, and health literacy scholarship. Sections~\ref{sec:limitations} and~\ref{sec:conclusion} address limitations and conclude.

%% file: chapter/2_related_work.tex
\section{Related Work}
\label{sec:related_work}

This section situates our investigation within the theoretical frameworks of ongoing conversations about mental health technology design, examining how HCI has approached digital mental health interventions and where critical gaps remain, particularly regarding accessibility and inclusive design practices.

\subsection{Designing for Mental Health in HCI}
\label{subsec:mh_in_hci}

HCI research has increasingly engaged with mental health and wellness technologies, exploring how interactive systems can support well-being~\cite{coyleInteractionDesignEmotional2012, calvoDesignWellbeingTools2019}. This body of work spans diverse contexts: workplace stress reduction~\cite{kevin_chapman_mental_2019}, academic anxiety management~\cite{kellyItsMissingMuch2021}, social connectivity's effects on mental health~\cite{balcombeEvaluationUseDigital2023}, and challenges faced by marginalized groups~\cite{ayobiDigitalMentalHealth2022}. Researchers have investigated emerging interaction paradigms including LLM-based chatbots for cognitive therapy~\cite{guMentalBlendEnhancingOnline2024, liAutomaticEvaluationMental2024}, VR-mediated exposure therapy~\cite{baghaeiTimeGetPersonal2020, feinbergZenVRDesignEvaluation2022, meinlschmidtMentalHealthMetaverse2023}, and smartphone-based mood tracking~\cite{alslaityInsightsLongitudinalEvaluation2022, hoeferVisualizingUncertaintyMultiSource2022}.

These technological innovations promise improved availability and engagement in mental health care~\cite{bowie-dabreoUserPerspectivesEthical2022, bunyiAccessibilityDigitalMental2021}, alongside opportunities for personalized goal-setting~\cite{zhang_designing_2024}. Yet critical concerns have emerged regarding privacy~\cite{hassanUnveilingPrivacyMeasures2023}, algorithmic bias~\cite{kangThisAppSaid2024}, therapeutic effectiveness~\cite{khooThatsKindSuspicious2024}, and power dynamics between users and systems~\cite{oguamanamIntersectionalLookUse2023, robledoyamamotoTherapyTeletherapyRelocating2021}. Notably, HCI's exploration of these technologies is limited, especially their effectiveness and accessibility for diverse populations, particularly people with disabilities~\cite{bunyiAccessibilityDigitalMental2021, sienDesigningInclusivityAccessibility2023}. Our work addresses this gap by centering blind users' experiences and design needs.

\subsubsection{The Evolution of Digital Mental Health Tracking Services}
\label{subsubsec:evolution_of_dmh}

Digital mental health (DMH) tracking services represent a particularly significant category of mental health technology, offering sustained engagement through mood monitoring~\cite{barryMHealthMaternalMental2017}, guided meditation and mindfulness practices~\cite{daudenroquetEvaluatingMindfulnessMeditation2018, markumDigitalTechnologyMeditative2020}, and peer support communities~\cite{olearySuddenlyWeGot2018}. These services have evolved from simple self-help applications to sophisticated AI-driven platforms offering personalized interventions~\cite{ausmanArtificialIntelligencesImpact2019, cabreraEthicalDilemmasMental2023}. Proponents argue DMH services reduce barriers to accessing support~\cite{adamsAvailabilityAccessibilityMental2024, lattieOverviewRecommendationsMore2022}, provide timely assistance~\cite{bowie-dabreoUserPerspectivesEthical2022, calvoPositiveComputingResearch2017}, and enable continuous mental health monitoring~\cite{kornfieldEnergyFiniteResource2020, nepalCurrentPracticesMental2021}.

However, their effectiveness compared to traditional in-person interventions remains contested~\cite{ausmanArtificialIntelligencesImpact2019, robledoyamamotoTherapyTeletherapyRelocating2021}. Prior work has documented challenges with user engagement, sustained adherence, and potential adverse effects. For instance, studies reveal financial inaccessibility among prenatal Black women of lower socioeconomic status~\cite{oguamanamIntersectionalLookUse2023} and practical burdens associated with carrying additional devices daily~\cite{sheikhWearableEnvironmentalSmartphoneBased2021}. These findings underscore that technology design alone cannot address mental health disparities; social, economic, and infrastructural factors profoundly shape who can benefit from DMH services and how.

\subsubsection{eHealth Literacy Framework}
\label{subsubsec:ehealth-framework}

Understanding varied user experiences with DMH services requires examining the competencies a user must bring to digital health engagement. Norman and Skinner's eHealth Literacy Framework~\cite{norman_ehealth_2006}, developed in 2006 and extended by \citet{norgaardEhealthLiteracyFramework2015} in 2015, provides one of the most widely cited analytical lenses for this purpose. The framework emerged from a gap in the literature: traditional health literacy models described what users needed to understand print health information~\cite{nutbeamHealthLiteracyPublic2000}, but did not account for the additional competencies required to locate, evaluate, and act on health information in digital environments. Norman and Skinner proposed that effective eHealth engagement requires the integration of six literacies:
 
\begin{itemize}
  \item \textbf{Traditional literacy and numeracy}: the ability to read, write, and work with quantitative information, which underpins all other literacies.
  \item \textbf{Health literacy}: understanding health concepts, medical terminology, and the functioning of healthcare systems, as originally articulated by \citet{nutbeamHealthLiteracyPublic2000}.
  \item \textbf{Information literacy}: locating relevant information across diverse sources and applying it to a specific problem.
  \item \textbf{Scientific literacy}: comprehending scientific principles, research methodology, and the nature of evidence.
  \item \textbf{Media literacy}: critically evaluating the provenance, framing, and trustworthiness of mediated content.
  \item \textbf{Computer literacy}: operating digital systems, including software, hardware, and networked environments.
\end{itemize}
 
Subsequent applications of the framework in digital health research have used it in three main ways: (1) as a tool for assessing user readiness to adopt eHealth interventions, (2) as a design heuristic for identifying which literacies a given technology implicitly demands, and (3) as a policy lens for understanding differential access to digital health. The framework has been particularly productive for research on underserved populations, where disparities in any one literacy can compound disparities in the others.
 
Applying this lens to DMH tracking services raises questions that are particularly acute for blind users. DMH services rely heavily on visual information presentation (charts, calendars, color-coded mood entries) and on visual navigation (tappable icons, layout-based information hierarchy). A user who cannot see the screen must use assistive technology: most commonly a screen reader, to mediate between themselves and the interface. The eHealth Literacy framework would describe this mediation as drawing on \textit{computer literacy}, but as we will argue in Section~\ref{subsubsec:literacies_for_dmh}, this characterization undersells what blind users are actually doing and overstates the degree to which the problem can be solved by improving users' competencies.
 
We adopt the eHealth Literacy framework in this paper for two reasons. First, it is the dominant model in digital health scholarship, and our findings need to be legible to that literature. Second, its six-literacy structure provides a useful decomposition that lets us locate barriers precisely: even, as we will show, when that decomposition requires extension.

When selecting our framework, we considered multiple options, such as digital well-being, given mental health's close relationship to well-being, and frameworks centered on agency or privacy, given RQ3's focus on data agency. Well-being-centered and agency-centered frameworks describe desirable outcomes: what a good relationship between a user and a digital health technology looks like. They do not, however, supply the analytic distinction our data demanded: whether a breakdown occurs because a user lacks a competency or because the design prevents an existing competency from being exercised. The eHealth Literacy framework's six-literacy decomposition allows us to locate precisely where breakdowns occur, and our literacy-gap versus access-barrier extension (Section~\ref{subsubsec:literacies_for_dmh}) is itself an intervention into that framework, which in its original formulation conflates the two conditions. In other words, we chose eHealth Literacy not because it is complete, but because it is the dominant lens whose incompleteness our data allows us to correct.

\subsubsection{Literacy vs.\ Access Barriers: A Conceptual Distinction.}
\label{subsubsec:literacies_for_dmh}

\input{figure/figures-in-text/ehealth_literacy_framework}

A recurring complication in applying the eHealth Literacy framework to accessibility research is that the framework treats literacies as competencies that individuals possess. Under this framing, when a user cannot accomplish a task, the analytic move is to locate a literacy gap and recommend targeted skill-building. But this framing can misrepresent what is actually happening when blind users fail to complete tasks in DMH services. Many of our participants were sophisticated users of assistive technology, well-versed in digital navigation, and capable of critically evaluating digital health information. Their challenge was not that they lacked computer or information literacy; it was that the services they tried to use were designed in ways that prevented them from applying their literacy to the task at hand.
 
We therefore distinguish two kinds of failure condition that the eHealth Literacy framework, in its original formulation, tends to collapse:
 
\begin{itemize}
  \item A \textbf{literacy gap} exists when a user lacks a competency the task requires: for example, a user who has never encountered DMH services may have an information literacy gap regarding how to evaluate which service fits their needs.
  \item An \textbf{access barrier} exists when a user possesses the relevant competency, but the design of the service prevents them from applying it: for example, an unlabeled button on a screen-reader-incompatible interface, or a paywall that hides accessibility information until after purchase.
\end{itemize}
 
This distinction carries multiple distinct values. First, it has different implications for intervention: literacy gaps can be addressed through user-facing education and support, while access barriers must be addressed through changes to the service itself. Second, it clarifies responsibility: access barriers are properly understood as design failures, not user failures, and the language of "literacy" can obscure that. Third, it makes visible a pattern that prior applications of the framework have struggled to account for: the phenomenon of users who are, by any reasonable measure, digitally literate, but who nevertheless cannot use the services in question.
 
We use this distinction throughout Sections~\ref{sec:findings} and~\ref{sec:discussion}; where our participants lacked a competency, we describe the situation as a literacy gap. Where they possessed the competency but could not exercise it, we describe the situation as an access barrier. Readers familiar with critical disability studies will recognize this distinction as aligned with the social model of disability~\cite{sharifShouldSayDisabled2022, spielNothingUsUs2020}, which locates disability in environments and institutions rather than in individual bodies.
 
\subsection{Accessible Mental Health Tracking for Blind Users}
\label{subsec:access_and_mh_for_blind_users}

The intersection of accessibility and mental health technology represents an emerging yet understudied area in HCI and related fields. While DMH services hold potential to democratize access to mental health support, their accessibility for disabled users, particularly the blind community, has limited understanding~\cite{khan_sighted_2025, kohdaMentalHealthStatus2019}. This knowledge gap is especially concerning given that blind individuals often face unique mental health challenges, including higher rates of depression and anxiety compared to the general population~\cite{kohdaMentalHealthStatus2019, richardsonUnderutilizationMentalHealth2024}.

Recent research has begun exploring blind individuals' needs and preferences in personal health tracking contexts~\cite{leeIdentifyAdaptPersist2024, millerSelfMonitoringPhysicalActivity2022, choi_exploring_2024}. These studies emphasize the importance of adaptive interfaces~\cite{hoeferVisualizingUncertaintyMultiSource2022, slovakHCIContributionsMental2024}, audio feedback mechanisms~\cite{chapmanSociotechnicalConsiderationsAccessibility2024, rectorExploringOpportunitiesChallenges2015}, and haptic interactions for conveying emotional information~\cite{rectorExploringOpportunitiesChallenges2015, soler-dominguezARCADIAGamifiedMixed2024}. However, systematic investigation of current DMH platforms' effectiveness for blind users remains limited. 

Khan et al.'s recent work~\cite{khan_sighted_2025} represents an important starting point, partnering with blind advocacy organizations to illuminate initial needs for DMH tracking services among blind users. Yet broader questions persist: How do blind individuals currently navigate existing DMH platforms? What workarounds have they developed? Which design patterns facilitate or hinder their mental health self-care practices? What aspirations do they hold for future DMH services?

\subsubsection{Design Opportunities at the Intersection}
\label{subsubsec:design_opportunities}

Understanding DMH tracking tool usage among blind individuals offers multiple design research opportunities. First, it helps identify specific accessibility gaps in current DMH service design~\cite{calvoComputingMentalHealth2016, calvoDesignWellbeingTools2019, lattieOverviewRecommendationsMore2022}, moving beyond generic accessibility guidelines to uncover context-specific barriers in mental health technology. Second, it reveals how technology might enable customized interventions that respect blind users' expertise and preferences~\cite{millerSelfMonitoringPhysicalActivity2022, pandeyMentalHealthEvaluation2023}. Third, it contributes to broader accessibility research by demonstrating how designing for disability can generate innovative interaction paradigms~\cite{sienCodesigningMentalHealth2023, slovakHCIContributionsMental2024}.

Despite progress in HCI research on mental health and personal health tracking, critical examination of these technologies' accessibility and usability for blind users remains nascent. Our research addresses this gap by systematically investigating blind individuals' experiences with DMH tracking services—their current usage patterns, adoption barriers, and design aspirations. Through this investigation, we aim to chart pathways toward more inclusive DMH solutions that center blind users' needs, practices, and expertise in the design process itself.

%% file: figure/figures-in-text/ehealth_literacy_framework.tex
\begin{figure*}[ht]
    \centering
        \includegraphics[width=0.4\textwidth]{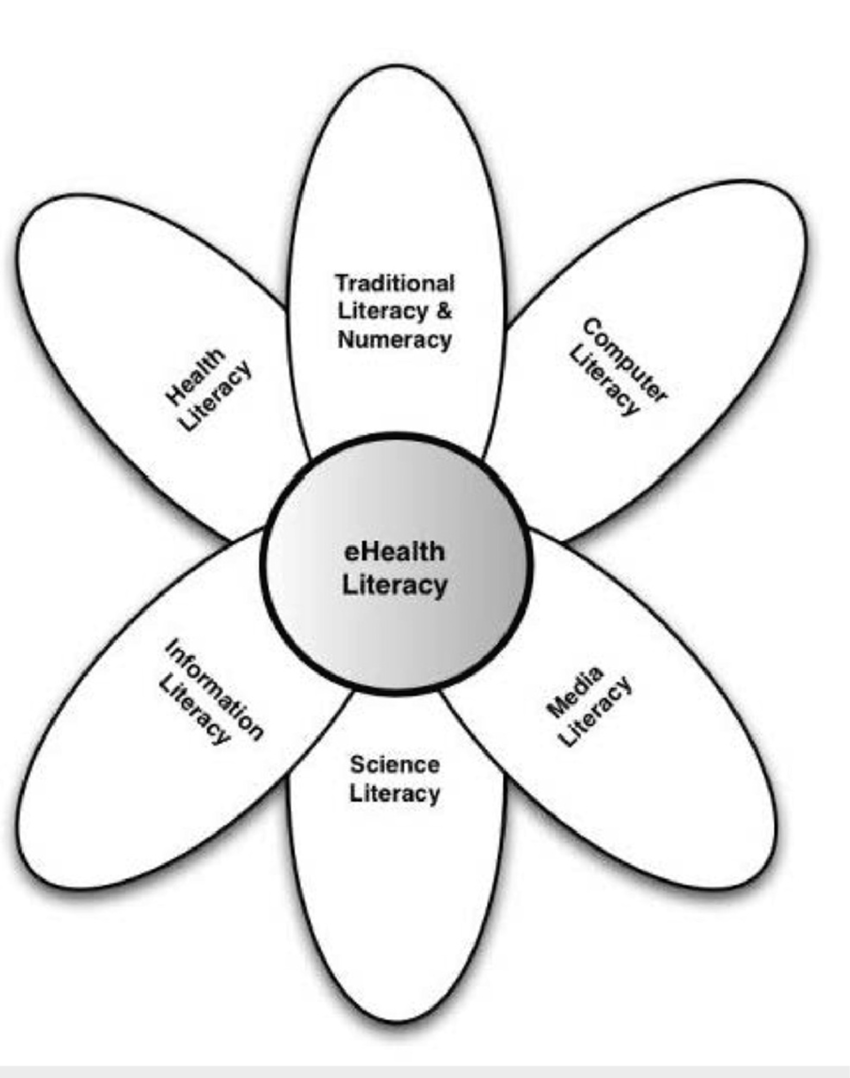}
        \caption{Norman and Skinner's eHealth Literacy framework~\cite{norman_ehealth_2006, norgaardEhealthLiteracyFramework2015}.}
        \Description{A flower-shaped diagram with eHealth Literacy at the center, surrounded by six petals representing its component literacies: Health Literacy, Traditional Literacy and Numeracy, Computer Literacy, Media Literacy, Science Literacy, and Information Literacy.}
    \label{fig:ehealth-literacy-framework}   
\end{figure*}

%% file: chapter/3_methods.tex
\section{Methods}
\label{sec:methods}
\begin{figure*}[ht]
    \centering
    \includegraphics[width=\textwidth]{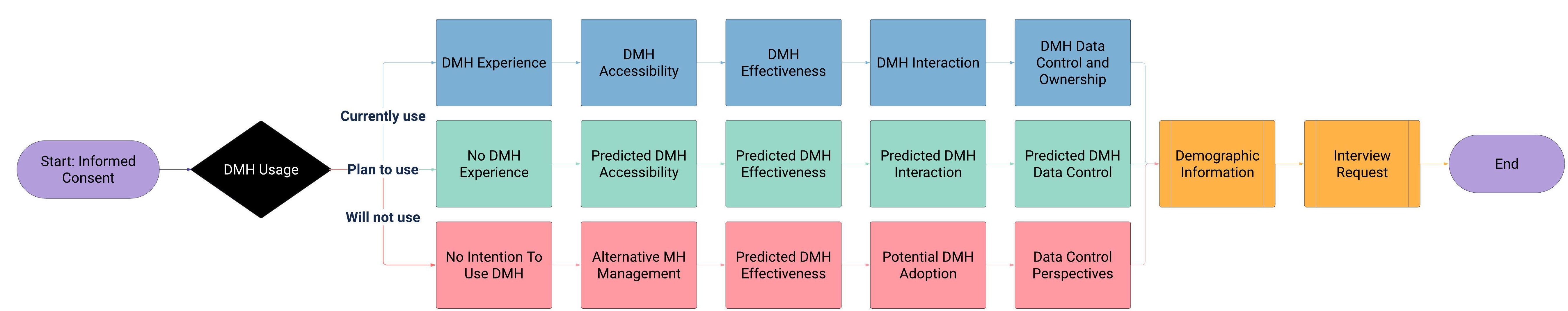}
    \Description[DMH Survey Flow Diagram]{Flow chart depicting the three tracks that survey respondents could have taken. The diagram starts with 'Informed Consent' and branches into three paths based on DMH Usage: 'Currently use', 'Plan to use', and 'Will not use'. Each path progresses through stages of Experience/No Experience, Accessibility, Effectiveness, Interaction, and Data Control. The 'Currently use' and 'Plan to use' paths converge at 'DMH Data Control and Ownership' and 'Predicted DMH Data Control ', respectively. Before ending, all paths lead to 'Demographic Information' and 'Interview Request'. The chart uses color coding: blue for current users, green for planned users, and maroon for non-users.}
    \caption{Flow diagram displaying the survey's flow. Participants received varying questions depending on their response to the survey's ``DMH Usage'' portion. Upon completing the DMH-related questions, all participants had the chance to complete the same demographic and interview request questions.}
    \label{fig:survey-flow}
\end{figure*}

To develop a nuanced understanding of the blind community's DMH tracking service use and broader MH technology needs, we designed an explanatory sequential mixed-methods study~\cite{creswell_designing_2017} that employed a quantitative phase (cross-sectional observational survey) to establish some breadth on DMH tracking service usage, and then proceeded to a qualitative phase (semi-structured interviews) that aimed to gather depth on user experiences. 

\subsection{Positionality}
\label{subsec:positionality}

To establish methodological transparency, we begin by acknowledging our positionality as researchers and its influence on our study design and analytical framework. The research team comprises two blind authors: the first author is low-vision, while the second author is blind. This experiential diversity within the research team enhanced our methodological approach by incorporating multiple perspectives on blindness, thereby strengthening our capacity to comprehensively examine how varying degrees of blindness impact users' engagement with DMH tracking services. Our distinct lived experiences informed both the conceptual framework and interpretative lens of this study, aligning with established practices in disability studies research that value disabled researchers' positionality as a methodological asset~\cite{mankoffDisabilityStudiesSource2010, sharifShouldSayDisabled2022}.

\subsection{Participant Recruitment}
\label{subsec:recruitment}

To build a comprehensive sense of the challenges faced by the blind community with DMH tracking services, we ensured that our study captured a holistic view of accessibility challenges and opportunities within this domain. Thus, our inclusion criteria included the following:
\begin{enumerate}
    \item The participant \textbf{must be 18 years of age or older}.
    \item The participant\textbf{ must identify as legally blind}.

    \item The participant \textbf{must be physically located in the United States (U.S)}.
\end{enumerate}

The participant pool was limited to the United States for this initial investigation to eliminate international payment processing complications and streamline survey timing and deployment.

We collaborated with several blind advocacy organizations within the U.S, including the National Federation of the Blind (NFB)~\footnote{https://nfb.org/}, the American Council of the Blind (ACB)~\footnote{https://www.acb.org/}, the American Foundation for the Blind (AFB)~\footnote{https://www.afb.org/}, and the DO-IT mailing lists managed by the University of Washington~\footnote{https://www.washington.edu/doit/}. Our university's Institutional Review Board (IRB) reviewed this study and marked it exempt. Following the review and approval of our community partners, they distributed our study materials via their mailing lists. Survey deployment began in August 2024, and 124 responses have been collected to date; 93 complete responses meeting our inclusion criteria were retained for analysis. Participants were randomly selected for follow-up semi-structured interviews per their indication to be contacted again. All participants were contacted via their provided e-mail. Following completion of the entire interview, participants were compensated with a USD 20 Amazon e-gift card.

\subsection{Study Design} 
\label{subsec:study-design}


Several factors played a crucial role in choosing the methodological approach for this study. The primary challenge was ensuring sufficient breadth and depth in responses to accurately capture the current use of MH technology by the blind community. Moreover, achieving data saturation with this population can be challenging, as such a pool is not readily accessible~\cite{smith_asset_2018}. To capture both breadth and depth of insights into whether the blind community engages with DMH tracking services and to what extent, we decided that an explanatory sequential mixed-methods study design~\cite{creswell_designing_2017}was the optimal choice for our inquiry for its ability to examine broader trends in DMH usage, elaborate on these trends, and then integrate quantitative and qualitative findings to gain a more comprehensive understanding of our RQs. We designed a survey that took inspiration from two validated questionnaires: the Technology Acceptance Model (TAM)~\cite{davis_perceived_1989} and the Client Satisfaction Questionnaire (CSQ)~\cite{attkissonClientSatisfactionQuestionnaire1982}. To curate the list of DMH tracking services we would be asking about, we took inspiration from several existing frameworks and taxonomies for classifying DMH interventions~\cite{pinedaUpdatedTaxonomyDigital2023, chenHybridCareMental2024, crossDigitalCumulativeComplexity2024}. We then designed a semi-structured interview protocol that ensured alignment to our RQs while allowing for flexibility in participant responses. This study design also allowed us to investigate a historically well-defined problem: the need for diverse voices when designing DMH services~\cite{kohdaMentalHealthStatus2019, bunyiAccessibilityDigitalMental2021}.
We designed a cross-sectional observational survey on Qualtrics to gather insights on blind individuals' current use of DMH tracking services, including their typical usage patterns of these services, their interaction flow and information retrieval methods, and what features they found particularly useful. Figure~\ref{fig:survey-flow} provides a visual of the different completion pathways of our survey. 

\subsubsection{Survey Accessibility Measures.}
\label{subsubsec:survey-access}

Because our survey population was exclusively blind adults, we designed and tested the Qualtrics instrument for screen-reader use before deployment. Both authors, as screen reader and magnification users, piloted the full instrument with their own assistive technology before launch. Our accessibility measures included: (1) single-column layouts with a logical reading order, so that screen readers encounter questions and response options in the intended sequence; (2) native radio-button and checkbox items rather than sliders, drag-to-rank, or other custom widgets that are unreliable under screen readers; (3) no CAPTCHA and no timed questions; (4) descriptive field labels programmatically associated with their inputs; and (5) plain-language phrasing throughout. We document these measures both for transparency and as a reference for researchers designing accessible online surveys without immediate blind and low-vision collaborators in their networks.

\subsection{Instrument Design and Construct Mapping}
\label{subsec:instrument-design}

\subsubsection{Instrument Design and Grounding in TAM and CSQ.}
\label{sec:tamcsq}

\input{table/survey-mapping}

We drew on two validated instruments in developing our survey. The first, the Technology Acceptance Model (TAM)~\cite{davis_perceived_1989}, is the dominant theoretical model of user acceptance of information technologies and identifies perceived usefulness and perceived ease of use as the primary determinants of adoption intention. The second, the Client Satisfaction Questionnaire (CSQ)~\cite{attkissonClientSatisfactionQuestionnaire1982}, is a validated instrument for measuring satisfaction with health and social services. We used both as conceptual scaffolding rather than as validated scales to administer verbatim; our survey items were adapted to the DMH context and to the specific concerns of blind users, which neither instrument was designed to cover. Table~\ref{tab:instrument-map} maps each survey construct to its TAM or CSQ grounding.
 
We adapted item structures and response-scale conventions from the CSQ (particularly for items addressing service satisfaction and non-adoption reasons) but did not include the CSQ's validated items. This is a limitation insofar as our survey's psychometric properties cannot be directly compared to CSQ-based studies, but it was a necessary adaptation given that CSQ's items do not cover accessibility or the structural features of the DMH marketplace.

\subsection{Survey Analysis}
\label{subsec:survey-analysis}

We implemented a comprehensive analytical framework that combined 
rigorous statistical methods for closed-ended questions with systematic qualitative analysis of open-ended questions (survey materials are provided in Appendix, Section~\ref{sec:survey-instrument}). Our closed-ended response analysis used a frequentist lens, chosen for its ability to provide objective results without requiring prior assumptions~\cite{fornacon-woodUnderstandingDifferencesBayesian2022} - particularly crucial given the limited existing research on blind individuals’ experiences with DMH tracking services. We conducted descriptive statistics for all closed-ended items, including measures of central tendency and variability. To ensure statistical rigor, we employed the Kruskal-Wallis test~\cite{kruskal_use_1952} to examine differences in the perceived effectiveness of different service categories. We followed a systematic coding protocol for open-ended responses using ATLAS.ti 24 for Mac~\cite{ATLASTI_2024}. This process involved initial open coding, axial coding to identify relationships between concepts, and selective coding to map open-ended responses to the appropriate facet(s) within the eHealth literacy framework~\cite{norman_ehealth_2006, norgaardEhealthLiteracyFramework2015}. 

\subsection{Interview Analysis}
\label{subsec:interview-analysis}

The first author conducted all interviews remotely via Zoom. Before the interview, participants were electronically sent the informed consent document for review, which outlined study goals, data collection procedures, data analysis procedures, and the researchers' contact information. The first author obtained verbal confirmation from each participant of their familiarity with informed consent procedures before commencing the interview. All interviews were recorded and transcribed using OpenAI's local Whisper~\cite{radford_robust_2023}. Interviews were analyzed using ATLAS.ti 24 for Mac~\cite{ATLASTI_2024}. When considering approaches for interview analysis, the research team determined that \textit{inductive thematic analysis} (ITA) was the optimal approach for its emphasis on uncovering themes grounded in participant data~\cite{cooper_thematic_2012}. While we did not intentionally follow Braun and Clarke's approach to thematic analysis~\cite{braun_using_2006}, our analysis process closely resembled this approach's six core tenets. The first author independently engaged in this process, discussing findings on a weekly basis with the last author. Following final theme formation, the first author categorized each theme under the appropriate literacy type from Norman and Skinner's eHealth Literacy framework~\cite{norman_ehealth_2006, norgaardEhealthLiteracyFramework2015}.

%% file: table/survey-mapping.tex
\begin{table*}[ht]
\centering
\caption{Mapping of survey constructs to TAM and CSQ theoretical grounding. We adapted item structures and response-scale conventions from these instruments; we did not administer either in its validated form.}
\label{tab:instrument-map}
\small
\begin{tabularx}{\textwidth}{@{}p{4.2cm} p{3.6cm} X@{}}
\toprule
\textbf{Survey construct} & \textbf{Source} & \textbf{Adaptation For Study Context} \\
\midrule
Perceived helpfulness (T1/T2.6) &
TAM: Perceived Usefulness &
Adapted to mental-health-specific outcomes rather than generic task performance. \\
Effectiveness rating (T1/T2.13) &
TAM (Perceived Usefulness) + CSQ (service effectiveness) &
Combined item capturing both perceived utility and realized service outcomes. \\
Accessibility rating (T1/T2.8) &
N/A &
Captured the degree to which the service works with the user's assistive technology. \\
Progress-tracking \newline difficulty (T1/T2.15) &
TAM: Perceived Ease of Use &
Specialized to the data-interpretation subtask central to DMH tracking. \\
Data control item battery (T1/T2.21--22) &
CSQ: satisfaction with service handling &
Extended to cover privacy, agency, and transparency dimensions not in CSQ. \\
Reasons for non-adoption (T3.1) &
CSQ: service utilization predictors &
Adapted from CSQ item structure but populated with accessibility- and cost-specific options. \\
Prerequisite features for adoption (T1/T2.18, T3.8) &
TAM: behavioral intention antecedents &
Reframed to surface accessibility prerequisites as adoption conditions. \\
\bottomrule
\end{tabularx}
\end{table*}

%% file: chapter/4_findings.tex
\section{Findings}
\label{sec:findings}

\input{figure/figures-in-text/q10_53_reasons_services_relationship}


We present our findings organized by research question. Within each RQ, we first report quantitative survey results, then qualitative survey findings, then interview findings. We interpret findings through the eHealth Literacy framework~\cite{norman_ehealth_2006, norgaardEhealthLiteracyFramework2015}, distinguishing literacy gaps from access barriers (Section~\ref{subsubsec:literacies_for_dmh}).

\subsection{Participants}
\label{subsec:interview-participants}

Of the 93 responses we considered in our analysis, 82 participants (88.2\%) expressed interest in a follow-up interview. We randomly selected 10 participants for interviews to ensure fairness in participant selection. The first author interviewed all 10 participants, conducting interviews remotely via Zoom at a time of the participant's choosing. The average length of each interview was about 31 minutes ($s$ = 9.4 minutes).

Ten interviews were sufficient for this study for two reasons. First, in an explanatory sequential design, the qualitative phase's role is to explain and deepen quantitative results rather than to independently generalize; depth is prioritized over breadth \cite{creswell_designing_2017}. Second, our interviews reached thematic saturation: the later interviews surfaced no codes that had not already appeared in earlier ones, a sample size consistent with comparable ASSETS interview studies conducted with the blind and low-vision community.

Table~\ref{tab:interviewees} in Appendix~\ref{subsec:interview-participants} summarizes the demographic characteristics of interview participants, including age range, gender, visual acuity, and whether their blindness was congenital or acquired. Throughout the findings, survey respondents are identified as P\# and interview participants as N\#.

When asked to name the specific services they used or planned to use (item T1/T2.3), participants most often named Calm, Headspace, Fitbit, and Apple Health, alongside a long tail of category-specific services. All named services operate in the United States, consistent with our inclusion criteria.

\subsection{RQ1: Medical Tracking and Access to Therapeutic Techniques}

\subsubsection{Usage Concentrates in Audio-First Categories.}
\label{sec:rq1}
  
Although 60.2\% of our survey respondents had no prior experience with DMH tracking services, those with experience reported their usage concentrated in two categories: mindfulness and meditation services and sleep tracking services. Figure~\ref{fig:reasons-services-relationship} depicts the relationship between service category and primary reason for use; because many of its cells contain small counts, we treat it as descriptive context and let the qualitative findings below carry the interpretive weight.
 
The pattern of ratings points toward mindfulness/meditation and sleep tracking as the categories where blind users are currently deriving most value.

To examine whether participants perceived some service categories as more effective than others, we conducted a Kruskal--Wallis test on effectiveness ratings across the ten service categories. The test revealed no statistically significant difference across categories, $H(9) = 5.54$, $p = .785$, $\varepsilon^2 = .05$; effectiveness ratings were uniformly moderate-to-high, with category medians of 4 on the 5-point scale. Because our sample size was constrained by the reach of prospective participants rather than set by an a priori power analysis (Section~\ref{subsec:survey-analysis}), this null result should be read as an absence of evidence for category-level differences, not as evidence that no such differences exist. We accordingly do not claim that particular categories were rated more effective; instead, the survey's contribution lies in the accessibility and cost barriers that recur across categories, which we develop under RQ2. The lower engagement with therapy and counseling services is partially reflected by the broader accessibility deficits we describe in Section~\ref{sec:rq2}.

\subsubsection{Therapeutic Techniques, Medical Tracking, and Professional Use.}
 
Open-ended responses surfaced several reasons blind users gravitated toward these categories. Some framed use in terms of co-occurring health management: P29 described layering DMH use with broader medical
tracking:
 
\begin{quote}
    "They helped me keep track of my medications, various doctor's
    appointments. I also have an app that I [use] to help me with my
    anxiety, helps me track my sleep cycle through my Apple Watch as
    well\ldots send information if my doctor would like
    it\ldots can be forwarded to them\ldots and I've been able to reduce
    the psychiatric medication that I was once on." (P29)
\end{quote}
 
\noindent Others framed use in terms of specific therapeutic techniques they wanted to access; P70 described relying on guided meditation content: "\ldots I find it to be very relaxing, especially with the guided meditations, and the great music choices they have available! This is true before bed, in the middle of the afternoon, etc." (P70).
 
A theme that emerged specifically from the survey open-ended responses, and which we did not anticipate, was the professional use case: respondents with training in mental health counseling reported exploring DMH services to recommend to clients. P13 wrote: "I have a masters in clinical mental health counseling and have explored them to recommend to clients" (P13). This dual personal/professional relationship to DMH services suggests that the blind community's engagement with these services is more multi-faceted than a consumer-only framing captures.
 
\subsubsection{Soundscapes and Longitudinal Insight.}
 
Interview participants reinforced the survey's emphasis on mindfulness and progress tracking, but added two dimensions that the survey did not capture at depth: the desire for environmental soundscapes and narratives, and the desire for both active and passive progress monitoring. Soundscapes and ambient audio were described by multiple participants as DMH-adjacent tools that worked well because they did not require sighted engagement. N1 described using them both for themselves and for family members:
 
\begin{quote}
    "There's a thing called my noise and it's like a kind of sound generator meditation app. And I use that\ldots sometimes I will use that either when I can't sleep at night or I use it as like a white noise kind of calming noise sound to like if we're in a new place to help get my kid to sleep and me to sleep too. [We use it to] drown out other people['s] noises." (N1)
\end{quote}
 
\noindent This utility is partly accessibility-driven: audio-only tools work regardless of screen reader compatibility, but it also highlights how participants valued soundscapes as interventions in their own right, not merely as accessible fallbacks.
 
Participants also expressed strong interest in longitudinal monitoring. N9 wanted retrospective insight into their affective patterns:
 
\begin{quote}
    "And so I feel like I've always been interested in like knowing different times or if I'm stressed or something like what how my mood lines up with that\ldots I'm always curious like I guess just like the data of\ldots just seeing\ldots if certain times of the year or certain times of the month or a week are more stressful than others for like mood tracking and things like that." (N9)
\end{quote}
 
\noindent N2 imagined a more synthetic, cross-modal version of this:
 
\begin{quote}
    "\ldots there needs to be like a seamless understandable, concise way of reading the data\ldots whether that's me telling Siri\ldots hey how has my way been over the past week and\ldots that would be nice if like it could correlate its other things where not only am I recording weight but I'm recording what I ate." (N2)
\end{quote}
 
\noindent Both participants asked for the same thing: a DMH service that produces narrative, longitudinal insight from the data it already collects. However, neither was describing a feature they had experienced. Current services, in their accounts, capture the data without delivering the insight.

\subsubsection{Framework Interpretation.}
 
Under the eHealth Literacy framework, these findings highlight blind users' strong engagement with information literacy: they know what they want from a DMH service, combined with access barriers, not literacy gaps, at the level of service design. The category-level pattern suggests that accessible audio-first features succeed, and that failures concentrate in features that assume visual interpretation (charts, timelines, calendars).

\subsection{RQ2: Adoption Hinges on Accessibility That Can Be Evaluated}
\label{sec:rq2}
 
\subsubsection{Bimodal Accessibility Ratings and Adoption Prerequisites.}

\input{figure/figures-in-text/q12_14_23_likert_ratings}
 
Three closed-ended items captured adoption-relevant ratings: perceived accessibility with assistive technology (Figure~\ref{subfig:access}), difficulty of tracking progress (Figure~\ref{subfig:progress}), and perceived helpfulness (Figure~\ref{subfig:helpfulness}). Accessibility ratings were bimodal: respondents frequently rated DMH services as "somewhat accessible", but a substantial subset rated them "somewhat inaccessible" or "very inaccessible". When asked which accessibility features were prerequisite to adoption (Figure~\ref{subfig:prereq-features}), the most frequently selected were screen reader compatibility, simplified clutter-free design, accessible data visualization, and transparent security/privacy practices. The frequency with which transparent security/privacy practices appeared as a prerequisite alongside the more expected accessibility items also foreshadows RQ3 findings.

\subsubsection{Three Recurring Access Barriers: Compatibility, Navigability, and Constrained Choice.} 

Open-ended responses about accessibility clustered around three types of access barrier. The first was failures of screen reader compatibility: P8 wrote that existing services were:

\begin{quote}
    "\ldots mostly unhelpful because they are at times not accessible with a screen reader. Furthermore, they require me to go out of my way to\ldots set them up, and\ldots learn their UI/UX which at times is significantly different from what I am used to with my device." (P8)
\end{quote}

\noindent The second was unlabeled or unnavigable interface elements: P67 described apps "where [they] couldn't even get past the startup/welcome screens," and others where "creating accounts was accessible, but the rest of it was not" (P67). The third was constrained choice; P65 captured this structural problem directly:

\begin{quote}
    "I don't have a lot of [or] is any choices of what apps or services I use, while sighted people have their pick of the litter." (P65)
\end{quote}

\noindent This last point frames access not just as a property of any one service but as a property of the DMH landscape as a whole: blind users select from a much smaller list. Financial access was also a recurring theme, as noted by P75: \textit{"Most have a huge cost associated with them that doesn't seem worth it, and many have accessibility challenges or glitches sprinkled throughout them."} (P75) P75's joint framing of cost and accessibility, evaluated jointly, foreshadowed a structural pattern we later uncovered in follow-up interviews: blind users are asked to pay for services before they can tell whether those services are usable.

\subsubsection{Accessibility-Native Design, Personalization, and Community as Adoption Conditions.}

Interview participants deepened each of these themes and introduced three that the survey did not capture at depth: (1) accessibility-native rather than retrofitted design, (2) personalization as a precondition of sustained use, and (3) community connection as a feature blind users actively seek but are excluded from.
 
On accessibility-native design, N2 articulated a vision that explicitly rejected workaround-based engagement: \textit{"I want it to be straightforward, go in, click a button, it's done. I don't want my mental health app to give me mental barriers"} (N2). "Barriers" here was used by participants repeatedly to describe what accessibility retrofits fail to remove. N6 described the downstream cost of accessibility retrofits that fail:
 
\begin{quote}
    "I tried that app\ldots it teased me like I'm able to set up my username, password all that stuff, and then it didn't. I couldn't use [or] utilize all the features because it wasn't accessible." (N6)
\end{quote}
 
\noindent Personalization mattered in ways that, in our participants' accounts, went beyond aesthetic preference. N8 contrasted affirmation-based and discipline-based DMH interventions:

\begin{quote}
    "The difference between something that's affirmation-based versus something that you have to look for is that affirmations\ldots sort of automatically pop up on your\ldots device. Whereas the meditation thing is something that has to be there's much more discipline\ldots driven\ldots rooted in discipline and intentionality." (N8)
\end{quote}
 
\noindent The design implication is that DMH services need to support both push-based and pull-based interactions, and need to let users configure the balance. N7 asked for a closely related feature: quick access to frequently used content from the home screen, without additional navigation steps, framed explicitly as a MH-relevant design request: \textit{"I don't need to be frustrated when I'm trying to settle down and get to sleep"} (N7). Frustration, in this account, is not a UX problem in the ordinary sense; it is constrained by the very mental state the service is supposed to support.

Community connection also emerged as a recurring theme, highlighting the real-world, social ramifications that DMH inaccessibility can have on its users. N6 described exclusion from a women's wellness community app that her non-blind friends used:

\begin{quote}
    "my\ldots non-blind female friends have the app, they talk about a community that they can engage in, and I wanted to be part of that. But I couldn't because it's not accessible." (N6)
\end{quote}
 
\noindent This exclusion is doubly consequential in the DMH context: peer community support is one of the therapeutic mechanisms DMH services offer, and inaccessible community features therefore exclude blind users from an intervention's active ingredient, not just its interface. N5 described the workaround: seeking accessibility information from other blind users rather than from service providers: \textit{"I also kind of scope out some of the blind review groups or blind people to kind of find out their accessibility experiences"} (N5). The information pathway here is itself an access barrier: blind users must build their own evaluation infrastructure because service providers do not surface accessibility information in an actionable form.

\subsubsection{Framework Interpretation}
 
Applying our distinction between literacy gaps and access barriers (Section~\ref{subsubsec:literacies_for_dmh}), the adoption challenges our participants described are overwhelmingly access barriers rather than literacy gaps. Participants demonstrated information literacy (knowing how to evaluate services), computer literacy (skilled AT users), and health literacy (capable of articulating their MH goals). What they lacked was a marketplace that allowed them to exercise these literacies, a marketplace where accessibility is evaluable before purchase, where community features are accessible, and where personalization is substantive rather than cosmetic.

\subsection{RQ3: Data Agency Requires Accessible, Pre-Purchase Transparency}
\label{sec:rq3}
 
\subsubsection{Privacy Concerns Are Near-Universal Among Respondents.}
 
The survey captured data concerns at multiple points. In response to our question about privacy or security concerns, 85.2\% of the 88 respondents who answered the open-ended privacy item described at least one concern. When listing prerequisites for adoption, 26 respondents selected "transparent security/privacy practices": the fourth most frequently selected prerequisite. When asked how important data privacy and control was to them, among prospective users, 68.4\% rated it as "very important" or "extremely important."

\subsubsection{Confidentiality, Agency as a Right, and Transparency.}

Three concerns recurred across survey open-ends. The first was confidentiality, framed in terms both of unauthorized access and of MH-specific stigma. P10 focused on unauthorized access: \textit{"I do not like the idea of people who are not related to my medical care having access to sensitive information"} (P10). P7 framed the concern in terms of discrimination risk:
 
\begin{quote}
    "[Confidentiality is] important to me because there is still a stigma around mental health, chronic health, disability, etc. And I do not want my medical information shared with others without my consent because that information could be viewed negatively and used against me." (P7)
\end{quote}
 
\noindent The second concern was data agency as a civic principle; as P39 wrote: \textit{"I am a human and have the right to control/manage my own affairs just like every other individual on this earth whether or not they have a disability."} (P39) The third was transparency in data collection, handling, and storage; as per P12: \textit{"I want to know who is seeing my information, but at the end of the day I need the tools that I need so will share some things. Not really hiding anything"} (P12).

\subsubsection{Data Practices, Accountability, and the Cost of Evaluation.}

Interview participants elaborated each of these concerns and added two we develop here as distinct themes: the need for accountability when security failures occur, and the interaction between cost and accessibility in DMH adoption decisions.

On data practices, several participants described partial or incomplete familiarity with how DMH services handle their data. N5 described evaluating services by looking for visible security affordances rather than by reading privacy policies: \textit{"I like ones that have a little bit of security that's kind of obvious, you know...I don't delve so deeply into the privacy statements or things like that as I should probably."} (N5) This is an honest account of a pattern likely shared with many users across populations, but the implication for DMH design is that privacy must be communicated through clear affordances, not only through policy text, especially given that screen reader navigation of long privacy policies is itself cognitively taxing.

Participants also wanted accountability when data practices failed. N4 described what they looked for: \textit{"I would like them saying if it does happen this is what we will do to make up for it or like make good or this is how we will go about fixing the issue if it does happen"} (N4). This goes beyond disclosure toward a commitment about response, which few current DMH services provide.
 
The cost dimension of RQ3 turned out to be closely intertwined with RQ2 in our interview data. Participants described a structural problem: DMH services often place accessibility-relevant information behind a paywall, so blind users cannot evaluate whether a service will work for them without first paying. N3 articulated this directly:

\begin{quote}
    "if I were to pay for it\ldots would it work from an accessibility standpoint\ldots if I'm going to invest that time and money and energy into it like any other customer I want to know that it's going to work with my accessibility software\ldots but I also had the question of like if I were to upgrade would it be the same\ldots because I've seen apps\ldots where it works and then one day it stops working or only certain features work." (N3)
\end{quote}
 
\noindent This pattern, which we term paywall-gated accessibility evaluation, amounts to a financial tax on blind users' information gathering, and represents the many shapes and forms of barriers to entry experienced by blind users when using DMH tracking services.

\subsubsection{Framework Interpretation.}

These findings point most directly to information literacy and media literacy in the eHealth Literacy framework: participants are trying to locate, evaluate, and act on information about how their data is handled. But the access barriers here are again structural: the information needed to make informed adoption decisions is often not surfaced in accessible form, and in the paywall-gated case is not available at all prior to financial commitment. Data agency for blind users therefore cannot be achieved purely through user-facing education; it requires changes to how DMH services disclose their accessibility and privacy postures before purchase.

%% file: figure/figures-in-text/q10_53_reasons_services_relationship.tex
\begin{figure*}[t]
    \centering
        \includegraphics[width=\textwidth]{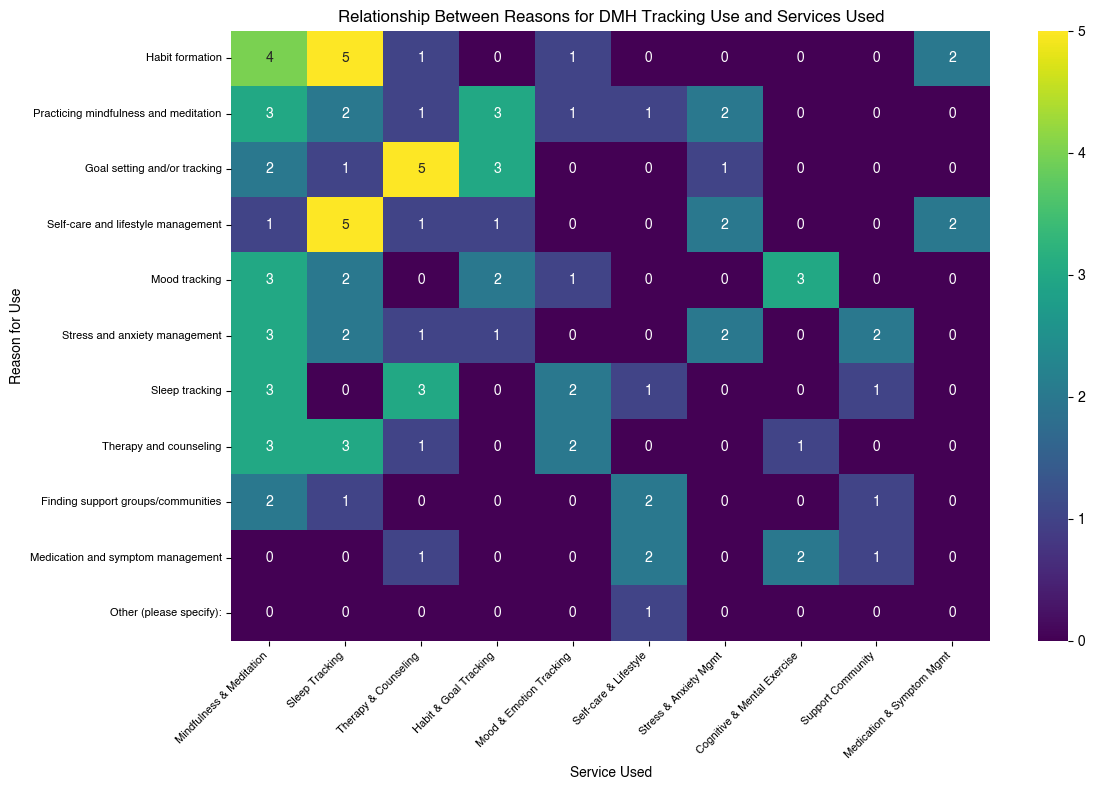}
        \caption{Heatmap illustrating the relationship between DMH tracking service usage frequency and reasons for use across the blind community.}
        \Description{A heatmap visualization showing the relationship between reasons for digital mental health (DMH) tracking use (y-axis) and services used (x-axis). The heatmap uses a color scale from dark purple (0) to bright yellow (5) to represent frequency or strength of relationships. The y-axis lists 11 reasons including habit formation, mindfulness practice, goal setting, self-care, mood tracking, stress management, sleep tracking, therapy support, finding support groups, medication management, and other. The x-axis shows 10 service categories including mindfulness \& meditation, sleep tracking, therapy \& counseling, and various management tools. The visualization reveals varying intensities of relationships between different reasons and services, with some notable clusters of higher values (yellows and greens) in certain intersections.}
    \label{fig:reasons-services-relationship}
\end{figure*}

%% file: figure/figures-in-text/q12_14_23_likert_ratings.tex
\begin{figure*}[t]
    \centering
    \begin{subfigure}[t]{0.33\textwidth}
        \centering
        \includegraphics[width=\textwidth]{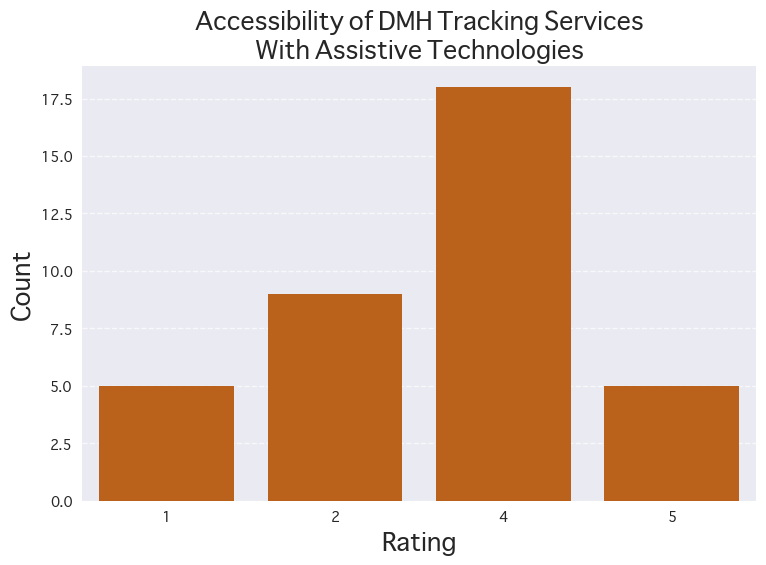}
        \caption{Distribution of participant ratings for accessibility. Ratings were measured on a 5-point Likert scale (1 = \textit{Very inaccessible}, 3 = \textit{Neither accessible nor inaccessible}, 5 = \textit{Very accessible}).}
        \Description{Bar chart displaying accessibility ratings of DMH Tracking Services with assistive technologies. X-axis shows ratings from 1-5, Y-axis shows count of responses. Distribution: Rating 1: 5 responses, Rating 2: 9 responses, Rating 4: 18 responses, Rating 5: 5 responses. Note: Rating 3 has no data.}
        \label{subfig:access}
    \end{subfigure}
    \hfill
    \begin{subfigure}[t]{0.3\textwidth}
        \centering
        \includegraphics[width=\textwidth]{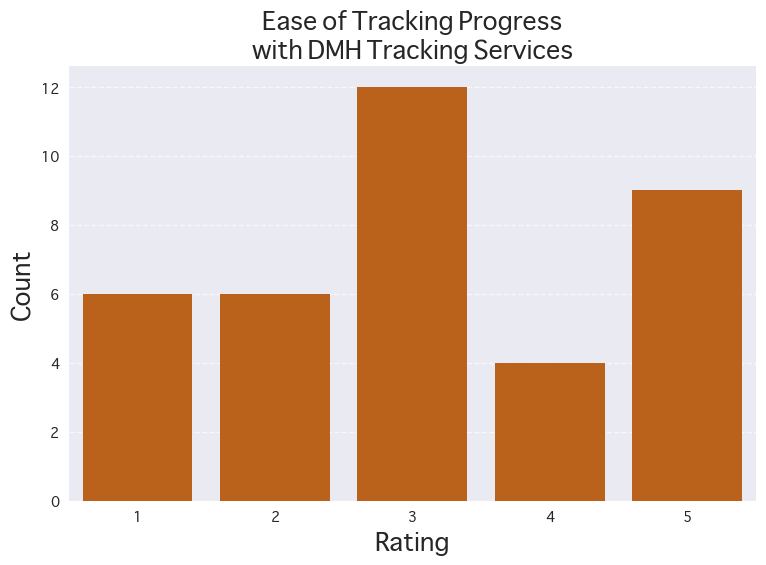}
        \caption{Distribution of participant responses to tracking progress difficulty. Ratings were measured on a 5-point Likert scale (1 = \textit{Very difficult}, 3 = \textit{Neither easy nor difficult}, 5 = \textit{Very easy}).}
        \Description{Bar chart showing distribution of ratings for ease of tracking progress with DMH tracking services. X-axis shows ratings from 1-5, Y-axis shows count of responses. Distribution: Rating 1: 6 responses, Rating 2: 6 responses, Rating 3: 12 responses, Rating 4: 4 responses, Rating 5: 9 responses.}
        \label{subfig:progress}
    \end{subfigure}
    \hfill
    \begin{subfigure}[t]{0.3\textwidth}
        \centering
        \includegraphics[width=\textwidth]{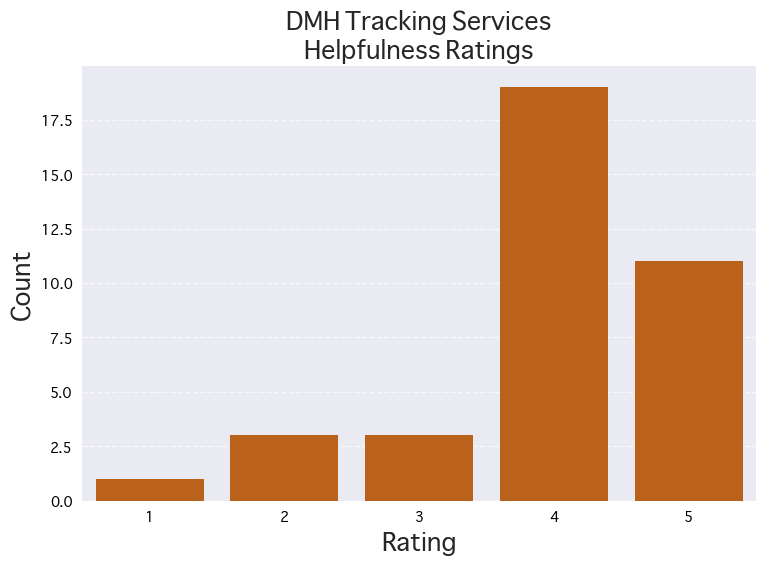}
        \caption{Distribution of participant ratings towards DMH helpfulness. Ratings were measured on a 5-point Likert scale (1 = \textit{Very unhelpful}, 3 = \textit{Neither helpful nor unhelpful}, 5 = \textit{Very helpful}).}
        \Description{Bar chart illustrating helpfulness ratings for DMH tracking services. X-axis shows ratings from 1-5, Y-axis shows count of responses. Distribution: Rating 1: 1 response, Rating 2: 3 responses, Rating 3: 3 responses, Rating 4: 19 responses, Rating 5: 11 responses.}
        \label{subfig:helpfulness}
    \end{subfigure}
    
    \caption{Participant ratings for DMH tracking services helpfulness and accessibility.}
    \Description{Three bar charts showing participant ratings for DMH tracking services' helpfulness, ease of tracking progress, and accessibility via assistive technologies arranged in a row.}
    \label{fig:dmh-distributions}
\end{figure*}

%% file: chapter/5_discussion.tex
\section{Discussion}
\label{sec:discussion}

Our findings reveal critical insights into how the blind community engages with DMH tracking services, illuminating both opportunities and barriers in current implementations. Building on these insights, we present potential design, development, and policy recommendations that not only address the specific needs of blind users but also offer transferable principles for improving accessibility and design across diverse populations. We hope to begin bridging the gap between empirical findings and practical implementation while considering broader implications for inclusive digital health technology.
 
\subsection{Digital Therapeutic Literacy in Blind Mental Health Management}
\label{sec:disc1}

Our finding that mindfulness/meditation and sleep tracking emerge as the top DMH categories among blind users diverges somewhat from patterns in the broader DMH literature, where mood tracking and therapy-oriented tools are often reported as the most-used categories \cite{ausmanArtificialIntelligencesImpact2019, cabreraEthicalDilemmasMental2023, kornfieldEnergyFiniteResource2020}. We offer two possibilities for this deviation. First, our participants may gravitate toward mindfulness and sleep tracking precisely because these categories are most likely to work with audio-first interaction: they are the categories where accessibility retrofits least undermine the core intervention. Second, the broader DMH literature's emphasis on mood tracking may reflect a research focus rather than a user preference; the blind community's preferences may be more representative of what users actually want when the choice is not constrained by what the research literature has studied.
 
The interview finding that blind users are drawn to and excluded from DMH community features extends Khan and Seo's earlier observation \cite{khan_sighted_2025} that blind users experience isolation when digital communities fail them. Here, we highlight that this is specifically consequential in the DMH context, where peer support is itself part of the therapeutic mechanism. When a women's wellness community is inaccessible, a blind user is not merely excluded from the service's interface; she is excluded from the intervention. This is a different kind of accessibility failure than the ones dominant in the accessibility literature \cite{chaudary_teleguidance-based_2021, choi_exploring_2024}, and it warrants a different kind of redress.
 
\subsection{Understanding DMH Service Adoption Through Information Pathways}
\label{sec:disc2}

The broader digital health acceptance literature treats adoption as driven primarily by perceived usefulness and ease of use \cite{davis_perceived_1989, mouloudj_factors_2023}. Our findings broadly confirm this, but point to an antecedent condition that the TAM does not adequately capture: the ability to evaluate the product before committing. For many of our participants, perceived usefulness cannot be formed before purchase, because accessibility, the precondition of perceived ease of use, is itself not visible until after payment. We refer to this as paywall-gated accessibility evaluation, and argue it deserves distinct treatment in DMH adoption theory. It is not that blind users have different adoption criteria; it is that they cannot apply the ordinary criteria until they have already paid the costs adoption was supposed to be contingent on. This extends prior work on digital health adoption barriers among marginalized populations \cite{mair_understanding_2025, oguamanamIntersectionalLookUse2023}. We hypothesize that paywall-gated accessibility evaluation likely operates for any user population whose relevant access needs are not surfaced by services before purchase.
 
Our community-exclusion finding also speaks to the tension between personalization and community support in DMH design \cite{zhangDesigningEmotionalWellbeing2021}. Prior work has generally treated these as dual goals; our findings suggest they compete in a specific way for blind users. When community features are inaccessible, blind users fall back on personalization; they build their own routines, curate their own soundscape collections, and seek accessibility information from peer networks outside the service. This is evidence of substantial user effort, not of user satisfaction. A DMH service that makes community features accessible would reduce the personalization burden on individual users.
 
\subsection{Technical Foundations for Accessible DMH Design}
\label{sec:disc3}

The DMH privacy literature has converged on a set of standard concerns: unauthorized data sharing, insufficient user control, opaque data practices \cite{hassanUnveilingPrivacyMeasures2023, khooThatsKindSuspicious2024, progga_womens_2025}. Our findings confirm these concerns are present among blind users, but they also surface additional dimensions. First, data agency for blind users depends on accessibility of the data-management interface, not only on the underlying policy. A service can have strong privacy policies and still leave blind users without meaningful agency if the privacy settings are themselves inaccessible. N5's account of evaluating security through visible affordances rather than reading policy text suggests that privacy communication must be designed for screen-reader-accessible quick assessment, not only for regulatory sufficiency. Second, accountability expectations among our participants went beyond disclosure toward commitment to response. Participants wanted services to commit, in advance, to what they would do if a breach occurred. This is a higher bar than current regulatory frameworks impose, and it is a reasonable one: blind users, who face elevated stigma risk around mental health data \cite{ausmanArtificialIntelligencesImpact2019}, have more at stake in data breaches than their non-blind peers.

\subsection{Implications for DMH Design, Policy, and Scholarship}
\label{sec:implications}

The three preceding discussion threads share a structure: in each, blind users possess the relevant literacies but are prevented from exercising them by decisions made upstream, in service design, in marketplace structure, or in how scholarship frames the problem. The implications below translate that structure into recommendations, and we map each implication to the specific finding it derives from so that the connection between evidence and recommendation is explicit.
 
\begin{itemize}
    \item \textbf{For DMH service designers: surface accessibility status before the paywall} (from the paywall-gated accessibility evaluation finding, Sections~\ref{sec:rq2} and~\ref{sec:disc2}). DMH services should expose, in their pre-purchase marketing material and at online listings, a structured accessibility statement that includes, at minimum, screen reader compatibility, keyboard navigability, and data visualization accessibility. This statement could be verifiable through a time-limited accessibility trial rather than a free-tier pilot that strips out the features blind users most need to evaluate.
    \item \textbf{For DMH service designers: treat community features as intervention-critical} (from the community-exclusion finding, Sections~\ref{sec:rq2} and~\ref{sec:disc1}). When community and peer-support features are part of a DMH service's therapeutic offering, their accessibility should be treated as intervention accessibility, not interface accessibility. This implies a different level of design investment and a different accountability standard.
    \item \textbf{For DMH service designers: make personalization substantive, not cosmetic} (from the push/pull and home-screen-access findings, Section~\ref{sec:rq2}). Participants asked for control over the balance between push-based interactions (affirmations that surface automatically) and pull-based interactions (content the user seeks out with intention), and for configurable shortcuts to frequently used content that bypass navigation entirely. These requests concern how the intervention reaches the user, not how the interface looks; personalization options limited to themes and layouts do not address them.
    \item \textbf{For DMH service providers and pricing teams: decouple evaluation from payment} (from the cost findings under RQ2 and RQ3). Cost recurred in our data both as a barrier in its own right and as the mechanism that makes paywall-gated accessibility evaluation costly to blind users specifically. Providers should ensure that the financial commitment a service asks for is proportionate to what a blind user can verify about the service beforehand, whether through accessibility trials, transparent listings, or refund policies that explicitly cover accessibility failures discovered after purchase.
    \item \textbf{For platform providers and regulators: require accessibility disclosure as a listing condition} (from the constrained-choice and information-pathway findings, Section~\ref{sec:rq2}). App marketplaces could require accessibility statements for DMH services as a condition of listing, following the model of other disclosure regimes (e.g., privacy labels on the iOS App Store). This would shift the cost of pre-purchase accessibility evaluation from blind users to service providers.
    \item \textbf{For accessibility and health literacy scholarship: diagnose before intervening} (from the framework interpretations across RQ1--RQ3 and Section~\ref{subsubsec:literacies_for_dmh}). The distinction between literacy gaps and access barriers deserves broader application across the eHealth Literacy framework's uses. Interventions that target literacy gaps are inadequate when the underlying problem is an access barrier, and vice versa; a prerequisite for any eHealth Literacy-informed intervention is a careful diagnosis of which of the two is operating in a given case.
\end{itemize}

%% file: chapter/6_limitations.tex
\section{Limitations}
\label{sec:limitations}

While our survey captured broad DMH tracking trends across platforms, this approach limited platform-specific insights. It lacked integration of other validated tools like the TUQ \cite{parmantoDevelopmentTelehealthUsability2016} for measuring telehealth usability and satisfaction. The dynamic DMH landscape suggests value in focused studies of specific services or demographics, particularly through longitudinal analysis. Moreover, as this was an observational survey and not an experimental survey, we cannot guarantee any generalizability, and further studies must be conducted to scale up to the broader blind and low-vision (BLV) community. Our cross-sectional methodology constrained temporal understanding of user engagement patterns and adaptation over time.

Our findings are scoped to the blind community and to mental health tracking specifically. We do not claim that the needs we document generalize to communities of people with other disabilities, whose interactions with DMH services may surface different barriers, nor to other health-tracking domains (e.g., physical activity, chronic condition management), where the balance of literacy gaps and access barriers may differ. We see both as important directions for future work: applying the literacy-gap versus access-barrier distinction to other disability communities would test the conceptual extension we propose, and applying it to other health-tracking domains would test whether paywall-gated accessibility evaluation is specific to the DMH marketplace or a general feature of digital health commerce.

%% file: chapter/7_conclusion.tex
\section{Conclusion}
\label{sec:conclusion}

We conducted an explanatory sequential mixed-methods study of blind users' engagement with digital mental health tracking services in the United States. Across a survey of 93 legally blind adults and follow-up interviews with 10 respondents, participants described a DMH marketplace in which their most-used categories: mindfulness, sleep tracking, and goal tracking: are the ones where accessibility retrofits least undermine the intervention, and in which adoption decisions are shaped by a structural feature we name \textit{paywall-gated accessibility evaluation}: accessibility information is typically not surfaced before purchase. Participants also described exclusion from DMH community features that, in the DMH context, are part of the therapeutic offering and not only of the interface.
 
Interpreting these findings through Norman and Skinner's eHealth Literacy framework, we argue that the framework needs extension for accessibility research: many of our participants' challenges reflected access barriers imposed by design, not literacy gaps held by users. We offered a distinction between the two that provides empirical insights across digital health accessibility research more broadly.
 
We advocate for several reforms in DMH tracking service design: surface accessibility before paywalls, treat community features as intervention accessibility, and hold services accountable to commitments about response when data practices fail. Designing DMH tracking services \textit{with} the blind community rather than \textit{for} them requires these structural changes, not accommodation at the margins.

%% file: chapter/9_acknowledgements.tex
\begin{acks}

We thank all community partners: the National Federation of the Blind (NFB), the American Council of the Blind (ACB), the American Foundation for the Blind (AFB), and the University of Washington's DO-IT Center for their assistance in distributing our study materials to their respective members and for their enthusiasm and support of this work. We would also like to thank our lab's fellow researchers for their feedback throughout the entire study. 

\end{acks}

%% file: chapter/10_appendix.tex
\clearpage

\appendix

\section{Appendix}
\label{sec:appendix}

\subsection{Survey Demographics}
\label{subsubsec:demographics}
\input{table/survey-demographics}

\subsection{Interview Demographics}
\label{appendix:interview-demographics}

\input{table/interview-demographics}

\subsection{Additional Survey Findings}
\label{subsec:survey-findings}

\input{figure/figures-in-text/q24_70_82_goals_prereqs_access}

\clearpage

\section{Survey Instrument}
\label{sec:survey-instrument}

\begin{enumerate}
    \item When was the last time you used a digital mental health tracking service?
    \begin{enumerate}
        \item Today
        \item This past week 
        \item This past month 
        \item This past year
        \item I have never used DMH tracking services.
    \end{enumerate}
\end{enumerate}
    \subsection*{For Current Users and Those Planning to Use Services}
    \begin{enumerate}
    \item[T1/T2.1] Which of the following digital mental health tracking services have you used before/do you plan to use?
    \begin{itemize}
        \item[\( \square \)] Cognitive and mental exercise services (e.g., Lumosity, Happify) 
        \item[\( \square \)] Habit and goal tracking services (e.g., Habitica, Strides) 
        \item[\( \square \)] Medication and symptom management (e.g., Medisafe, Bearable)  
        \item[\( \square \)] Mindfulness and meditation services (e.g., Headspace, Calm)
        \item[\( \square \)] Mood and emotion tracking services (e.g., Daylio, Moodfit)  
        \item[\( \square \)] Self-care and lifestyle management (e.g., Fabulous, Shine)  
        \item[\( \square \)] Sleep tracking services (e.g., Sleep Cycle, Fitbit) 
        \item[\( \square \)] Stress and anxiety management services (e.g., Sanvello, Pacifica, What's Up?) 
        \item[\( \square \)] Support community services (e.g., 7 Cups, NAMI)  
        \item[\( \square \)] Therapy and counseling services (e.g., MoodTools, Woebot)  
    \end{itemize}

    \item[T1/T2.2] Of the digital mental health tracking services that you have used/that you plan to use, when you have used them, how often did you use them/how often will you use them (see Table~\ref{tab:service-matrix})~\footnote{1 = Never, 2 = Rarely, 3 = Once a month, 4 = Several times a month, 5 = Once a week, 6 = Several times a week, 7 = Daily.}?
    \end{enumerate}
    
\begin{table*}[t]
    \centering
    \caption{Matrix of service types from which participants would indicate their frequency of use.}
    \label{tab:service-matrix}
    \begin{tabular}{|l|c|c|c|c|c|c|c|}
    \hline
    \textbf{Service} & \textbf{1} & \textbf{2} & \textbf{3} & \textbf{4} & \textbf{5} & \textbf{6} & \textbf{7} \\
    \hline
    Cognitive and mental exercise services & \(\bigcirc\) & \(\bigcirc\) & \(\bigcirc\) & \(\bigcirc\) & \(\bigcirc\) & \(\bigcirc\) & \(\bigcirc\) \\
    Habit and goal tracking services & \(\bigcirc\) & \(\bigcirc\) & \(\bigcirc\) & \(\bigcirc\) & \(\bigcirc\) & \(\bigcirc\) & \(\bigcirc\) \\
    Medication and symptom management & \(\bigcirc\) & \(\bigcirc\) & \(\bigcirc\) & \(\bigcirc\) & \(\bigcirc\) & \(\bigcirc\) & \(\bigcirc\) \\
    Mindfulness and meditation services & \(\bigcirc\) & \(\bigcirc\) & \(\bigcirc\) & \(\bigcirc\) & \(\bigcirc\) & \(\bigcirc\) & \(\bigcirc\) \\
    Mood and emotion tracking services & \(\bigcirc\) & \(\bigcirc\) & \(\bigcirc\) & \(\bigcirc\) & \(\bigcirc\) & \(\bigcirc\) & \(\bigcirc\) \\
    Self-care and lifestyle management & \(\bigcirc\) & \(\bigcirc\) & \(\bigcirc\) & \(\bigcirc\) & \(\bigcirc\) & \(\bigcirc\) & \(\bigcirc\) \\
    Sleep tracking services & \(\bigcirc\) & \(\bigcirc\) & \(\bigcirc\) & \(\bigcirc\) & \(\bigcirc\) & \(\bigcirc\) & \(\bigcirc\) \\
    Stress and anxiety management services & \(\bigcirc\) & \(\bigcirc\) & \(\bigcirc\) & \(\bigcirc\) & \(\bigcirc\) & \(\bigcirc\) & \(\bigcirc\) \\
    Support community services & \(\bigcirc\) & \(\bigcirc\) & \(\bigcirc\) & \(\bigcirc\) & \(\bigcirc\) & \(\bigcirc\) & \(\bigcirc\) \\
    Therapy and counseling services & \(\bigcirc\) & \(\bigcirc\) & \(\bigcirc\) & \(\bigcirc\) & \(\bigcirc\) & \(\bigcirc\) & \(\bigcirc\) \\
    \hline
    \end{tabular}
    \Description{Matrix of service types from which participants would indicate their frequency of use.}
\end{table*}

\begin{enumerate}
    \item[T1/T2.3] Please list which specific digital mental health tracking services you have used/plan to use.

    \item[T1/T2.4] What were the high-level reasons you decided to start using digital mental health tracking services?
    \begin{itemize}
        \item[\( \square \)] Finding support groups/communities
        \item[\( \square \)] Goal setting and/or tracking  
        \item[\( \square \)] Habit formation 
        \item[\( \square \)] Medication and symptom management 
        \item[\( \square \)] Mood tracking 
        \item[\( \square \)] Practicing mindfulness and meditation 
        \item[\( \square \)] Self-care and lifestyle management 
        \item[\( \square \)] Stress and anxiety management
        \item[\( \square \)] Sleep tracking  
        \item[\( \square \)] Therapy and counseling 
        \item[\( \square \)] Other (please specify): 
    \end{itemize}

    \item[T1/T2.5] If comfortable sharing, please elaborate on these reasons to help us understand why you decided to start using these services. 

    \item[T1/T2.6] To what extent have digital mental health tracking services been helpful/do you anticipate them being helpful in managing your mental health needs?
    \begin{itemize}
        \item[\(\bigcirc\)] Very unhelpful 
        \item[\(\bigcirc\)] Somewhat unhelpful
        \item[\(\bigcirc\)] Neither helpful nor unhelpful 
        \item[\(\bigcirc\)] Somewhat helpful
        \item[\(\bigcirc\)] Very helpful  
    \end{itemize}

    \item[T1/T2.7] What did you find/anticipate finding helpful or unhelpful about these services? 

    \item[T1/T2.8] How accessible are digital mental health tracking services for you when using tools like assistive technologies?
    \begin{itemize}
        \item[\(\bigcirc\)] Very inaccessible 
        \item[\(\bigcirc\)] Somewhat inaccessible
        \item[\(\bigcirc\)] Neither accessible nor inaccessible 
        \item[\(\bigcirc\)] Somewhat accessible
        \item[\(\bigcirc\)] Very accessible  
    \end{itemize}

    \item[T2.9] What alternative methods, if any, do you use to track your mental health?
    \begin{itemize}
        \item[\( \square \)] In-person therapy or counseling services 
        \item[\( \square \)] Journaling
        \item[\( \square \)] In-person support groups
        \item[\( \square \)] Manual tracking systems
        \item[\( \square \)] Medication and treatment diaries 
        \item[\( \square \)] Physical activity 
        \item[\( \square \)] Self-assessment questionnaires
        \item[\( \square \)] Spiritual and/or religious practices
        \item[\( \square \)] Other (please specify):
        \item[\( \square \)] I do not currently use any strategies to manage my mental health.  
    \end{itemize}

    \item[T1/T2.10] Which of the following accessibility issues have you encountered when using digital mental health tracking services?
    \begin{itemize}
        \item[\( \square \)] Confusing layout or navigation 
        \item[\( \square \)] Inaccessible graphs or visual data representations 
        \item[\( \square \)] Lack of screen reader support   
        \item[\( \square \)] Overly complicated security features   
        \item[\( \square \)] Poor color contrast   
        \item[\( \square \)] Timing out during data entry   
        \item[\( \square \)] Other (please specify):   
    \end{itemize}
    
    \item[T1/T2.11] How accessible are digital mental health tracking services for you \textbf{from a financial perspective}?
    \begin{itemize}
        \item[\(\bigcirc\)] Very inaccessible 
        \item[\(\bigcirc\)] Somewhat inaccessible
        \item[\(\bigcirc\)] Neither accessible nor inaccessible 
        \item[\(\bigcirc\)] Somewhat accessible
        \item[\(\bigcirc\)] Very accessible  
    \end{itemize}
    
    \item[T1/T2.12] Please describe your overall experiences with accessibility on digital mental health tracking platforms (technical, financial, etc.). 

    \item[T1/T2.13] From your experience, how effective have digital mental health tracking services been at meeting your mental health goals?
    \begin{itemize}
        \item[\(\bigcirc\)] Very ineffective 
        \item[\(\bigcirc\)] Somewhat ineffective
        \item[\(\bigcirc\)] Neither effective nor ineffective 
        \item[\(\bigcirc\)] Somewhat effective
        \item[\(\bigcirc\)] Very effective 
    \end{itemize}

    \item[T1/T2.14] Please describe what made these services ineffective or effective for you and why. 
    
    \item[T1/T2.15] How difficult is it for you to track your progress towards achieving mental health goals using digital mental health tracking platforms?
    \begin{itemize}
        \item[\(\bigcirc\)] Very difficult 
        \item[\(\bigcirc\)] Somewhat difficult
        \item[\(\bigcirc\)] Neither easy nor difficult 
        \item[\(\bigcirc\)] Somewhat easy
        \item[\(\bigcirc\)] Very easy 
    \end{itemize}
    
    \item[T1/T2.16] What assistive technologies, if any, do you typically use when interacting with digital mental health tracking services?
    \begin{itemize}
        \item[\( \square \)] Accessibility and/or usability barriers
        \item[\( \square \)] Braille display 
        \item[\( \square \)] Customized app/browser extensions or add-ons
        \item[\( \square \)] Keyboard navigation  
        \item[\( \square \)] Magnification software 
        \item[\( \square \)] Refreshable Braille keyboard 
        \item[\( \square \)] Screen customization (e.g., high contrast, text size) 
        \item[\( \square \)] Screen reader  
        \item[\( \square \)] Voice commands 
        \item[\( \square \)] Other (please specify): 
        \item[\( \square \)] I do not use any assistive technologies when interacting with these services. 
    \end{itemize}

    \item[T1/T2.17] What challenges, if any, do you encounter when trying to view your goal progress in a digital mental health tracking service?
    \begin{itemize}
        \item[\( \square \)] Accessing historical data  
        \item[\( \square \)] Exporting data in screen reader-friendly formats 
        \item[\( \square \)] Filling out security checks (like CAPTCHA)  
        \item[\( \square \)] Finding and using help guides 
        \item[\( \square \)] Interpreting data presented in visual formats 
        \item[\( \square \)] Navigating between different sections of the platform 
        \item[\( \square \)] Responding to time-limited messages or alerts 
        \item[\( \square \)] Understanding information that uses colors to show meaning 
        \item[\( \square \)] Other (please specify):  
        \item[\( \square \)] I do not face any challenges while using digital mental health tracking services. 
    \end{itemize}

    \item[T1/T2.18] What features or accommodations would need to be present for you to consider using a digital mental health tracking service?
    \begin{itemize}
        \item[\( \square \)] Accessible data visualization 
        \item[\( \square \)] Customizable layout or navigation 
        \item[\( \square \)] Multi-modal input (voice, text, touch) 
        \item[\( \square \)] Screen reader compatibility 
        \item[\( \square \)] Simplified, clutter-free design  
        \item[\( \square \)] Specialized accessibility support 
        \item[\( \square \)] Transparent security/privacy practices
        \item[\( \square \)] Other (please specify): 
    \end{itemize}
    
    \item[T1/T2.19] Could you please briefly elaborate on the challenges you mentioned or describe any other challenges you face with digital mental health tracking services? 

    \item[T1/T2.20] Could you please describe why you don't face any challenges with these services? 
    
    \item[T1/T2.21] Are there any features of the digital mental health tracking services that you have had experience with that have made you feel in control/might make you feel in control of your data on the service? If so, please briefly describe what these features are. If not, please write "N/A".
    
    \item[T1/T2.22] Do you have any privacy or security concerns about using digital mental health platforms? If so, please briefly describe them. If not, please briefly describe why that is the case.
\end{enumerate}
    \subsection*{For Those Not Planning to Use Services}
    \begin{enumerate}
    \item[T3.1] Please indicate why you do not plan on using digital mental health tracking services in the future or why you have stopped using them.
    \begin{itemize}
        \item[\( \square \)] Accessibility and/or usability barriers
        \item[\( \square \)] Cost of use 
        \item[\( \square \)] Lack of awareness about these services  
        \item[\( \square \)] Preference for traditional, in-person methods
        \item[\( \square \)] Security/privacy concerns 
        \item[\( \square \)] Skepticism about effectiveness
        \item[\( \square \)] Time constraints (not enough time to learn a new technology) 
        \item[\( \square \)] Other (please specify): 
    \end{itemize}

    \item[T3.2] If comfortable sharing, please describe your reasons below to help us better understand your reasons for not wanting to use digital mental health tracking services.

    \item[T3.3] What alternative methods, if any, do you use to track your mental health?
    \begin{itemize}
        \item[\( \square \)] In-person therapy or counseling services 
        \item[\( \square \)] Journaling
        \item[\( \square \)] In-person support groups
        \item[\( \square \)] Manual tracking systems
        \item[\( \square \)] Medication and treatment diaries 
        \item[\( \square \)] Physical activity 
        \item[\( \square \)] Self-assessment questionnaires
        \item[\( \square \)] Spiritual and/or religious practices
        \item[\( \square \)] Other (please specify):
        \item[\( \square \)] I do not currently use any strategies to manage my mental health.  
    \end{itemize}

        \item[T3.4] If comfortable sharing, please describe how you currently manage your mental health goals without using digital mental health tracking services.

    \item[T3.5] What strategies, if any, are you interested in trying to achieve your mental health goals?

    \item[T3.6] Have you had any negative experiences with other digital health services in general that have influenced your decision not to use digital mental health tracking services?
    \begin{itemize}
        \item[\(\bigcirc\)] Yes
        \item[\(\bigcirc\)] No
        \item[\(\bigcirc\)] I have not used other digital health services.
    \end{itemize}

    \item[T3.7] If comfortable sharing, could you briefly describe these experiences to help us better understand your experiences with digital mental health tracking services?

    \item[T3.8] What features or accommodations would need to be present for you to consider using a digital mental health tracking service?
    \begin{itemize}
        \item[\( \square \)] Accessible data visualization 
        \item[\( \square \)] Customizable layout or navigation 
        \item[\( \square \)] Multi-modal input (voice, text, touch) 
        \item[\( \square \)] Screen reader compatibility 
        \item[\( \square \)] Simplified, clutter-free design  
        \item[\( \square \)] Specialized accessibility support 
        \item[\( \square \)] Transparent security/privacy practices
        \item[\( \square \)] Other (please specify): 
    \end{itemize}

    \item[T3.9] Please explain why your selected features might influence your decision to use digital mental health tracking services.

    \item[T3.10] How important is data privacy and control to you when considering mental health management tools?
    \begin{itemize}
        \item[\(\bigcirc\)] Extremely important  
        \item[\(\bigcirc\)] Very Important
        \item[\(\bigcirc\)] Moderately important  
        \item[\(\bigcirc\)] Slightly important
        \item[\(\bigcirc\)] Not at all important  
    \end{itemize}

    \item[T3.11] Please describe why data privacy and control with mental health management tools is (or isn't) important to you.
\end{enumerate}
    \subsection*{Demographics}
    
    \begin{enumerate}
    \item[D.1] What is your age in years?
    \begin{itemize}
        \item[\(\bigcirc\)] 18-24
        \item[\(\bigcirc\)] 25-34 
        \item[\(\bigcirc\)] 35-44 
        \item[\(\bigcirc\)] 45-54 
        \item[\(\bigcirc\)] 55-64 
        \item[\(\bigcirc\)] 65 or older
    \end{itemize}

    \item[D.2] What is your gender?
    \begin{itemize}
        \item[\(\bigcirc\)] Woman
        \item[\(\bigcirc\)] Man
        \item[\(\bigcirc\)] Non-binary
        \item[\(\bigcirc\)] Prefer to self-describe
        \item[\(\bigcirc\)] Prefer not to say
    \end{itemize}

    \item[D.3] What is your level of visual acuity?
    \begin{itemize}
        \item[\(\bigcirc\)] Totally blind (no light or shape perception)  
        \item[\(\bigcirc\)] Legally blind, with both light and shape perception  
        \item[\(\bigcirc\)] Legally blind, with only light perception  
        \item[\(\bigcirc\)] Legally blind, with only shape perception  
        \item[\(\bigcirc\)] Legally blind, central vision loss  
        \item[\(\bigcirc\)] Legally blind, peripheral vision loss  
        \item[\(\bigcirc\)] Legally blind, tunnel vision  
        \item[\(\bigcirc\)] Legally blind, blurry vision  
        \item[\(\bigcirc\)] Legally blind, fluctuating vision  
        \item[\(\bigcirc\)] Legally blind, partial sight 
        \item[\(\bigcirc\)] Prefer to self-describe 
    \end{itemize}
    \end{enumerate}

    \subsection*{Follow-up Contact}
    \begin{enumerate}
    \item[F.1] Are you interested in being contacted about a paid follow-up interview with a graduate student researcher from the research team?
    \begin{itemize}
        \item[\(\bigcirc\)] No
        \item[\(\bigcirc\)] Yes
    \end{itemize}

    \item[F.2] Please enter your preferred email address so that we may contact you for a follow-up interview.

\end{enumerate}

%% file: table/survey-demographics.tex
\newlength{\tab}
\setlength{\tab}{0.1em}

\begin{table}[h]
     \caption{Demographic characteristics of survey respondents considered in final analysis (n=93).}
\begin{tabular}[t]{ll}
    \toprule
    \textbf{Demographic} & \textbf{n}\\
    \hline
    \textbf{Participants} & \textbf{93}\\
    \hline
    \rowcolor{blue!20} 
    \textbf{Gender} & \\
    \hspace{\tab} Woman & 64 \\
    \hspace{\tab} Man & 25 \\
    \hspace{\tab} Non-binary & 3 \\
    \hspace{\tab} Prefer not to say & 1 \\
    \midrule
    \rowcolor{blue!20}
    \textbf{Age} & \\
    \hspace{\tab} 35-44 & 26 \\
    \hspace{\tab} 25-34 & 19 \\
    \hspace{\tab} 55-64 & 17 \\
    \hspace{\tab} 45-54 & 16 \\
    \hspace{\tab} 65 or older & 7 \\
    \hspace{\tab} 18-24 & 7 \\
    \midrule
    \rowcolor{blue!20}
    \textbf{Visual Acuity} & \\
    \hspace{\tab} Totally blind (no light or shape perception) & 38 \\
    \hspace{\tab} Legally blind, with only light perception & 18 \\
    \hspace{\tab} Legally blind, partial sight & 8 \\
    \hspace{\tab} Prefer to self-describe & 7 \\
    \hspace{\tab} Legally blind, with both light and shape perception & 6  \\
    \hspace{\tab} Legally blind, peripheral vision loss & 6 \\
    \hspace{\tab} Legally blind, central vision loss & 4 \\
     \hspace{\tab} Legally blind, fluctuating vision & 3 \\
    \hspace{\tab} Legally blind, blurry vision & 1 \\
    \hspace{\tab} Legally blind, tunnel vision & 1 \\
    \hspace{\tab} Legally blind, with only shape perception & 1 \\
    \bottomrule
    \end{tabular}
    \Description{Table showing demographic characteristics (gender, age, and visual acuity) of DMH tracking service users throughout the blind community (n=93). Women comprised the majority of participants' gender (64), followed by men (25), non-binary (3) and prefer not to say (1). The 35-44 age range was the most represented (26), followed by 25-34 (19), 55-64 (17), 45-54 (16), 65 or older (7), and 18-24 (7). Totally blind (no light or shape perception) was the most represented visual acuity (38), followed by legally blind - with only light perception (18), Legally blind -  partial sight (8), Prefer to self-describe (7), Legally blind - with both light and shape perception (6), Legally blind, peripheral vision loss (6), Legally blind - central vision loss (4), Legally blind - fluctuating vision (3), Legally blind - blurry vision (1), Legally blind - tunnel vision (1), and Legally blind - with only shape perception (1).}
    \label{tab:survey-demographics}
    \hfill
\end{table}

%% file: table/interview-demographics.tex
\begin{table}[h]
    \caption{Interview participant demographics ($n = 10$). Visual acuity categories follow the survey's self-description options (Appendix~\ref{sec:survey-instrument}, item D.3).}
    \label{tab:interviewees}
    \begin{tabular}{lllll}
        \toprule
        ID & Age & Gender & Visual acuity \\
        \midrule
        N1  & 25-34 & Man & Legally blind \\
        N2  & 18-24 & Man & Legally blind \\
        N3  & 25-34 & Man & Legally blind, central vision loss \\
        N4  & 55-64 & Woman & Legally blind \\
        N5  & 35-44 & Woman & Totally blind (no light or shape perception) \\
        N6  & 25-34 & Woman & Legally blind \\
        N7  & 25-34 & Woman & Legally blind \\
        N8  & 25-34 & Woman & Legally blind \\
        N9  & 45-54 & Man & Totally blind (no light or shape perception) \\
        N10 & 25-54 & Woman & Legally blind \\
        \bottomrule
    \end{tabular}
\end{table}

%% file: figure/figures-in-text/q24_70_82_goals_prereqs_access.tex
\begin{figure}[h!b]
    \centering
        \begin{subfigure}[b]{0.8\textwidth}
        \centering
        \includegraphics[width=\textwidth]{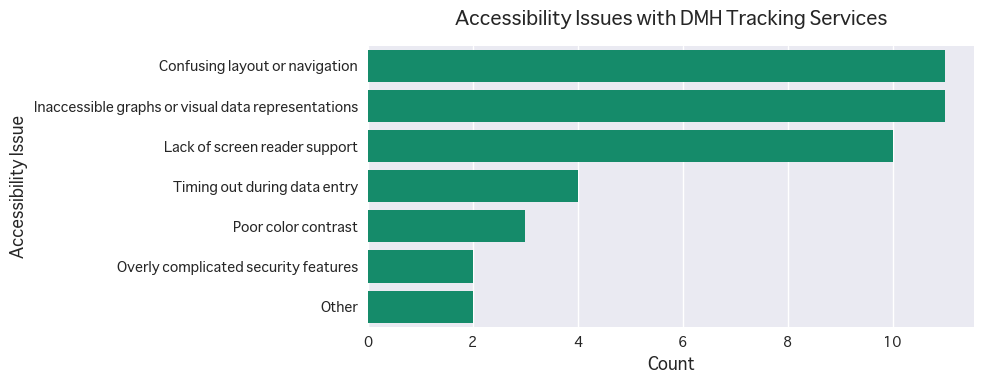}
        \caption{Reported access challenges when using DMH tracking services.}
        \Description{Horizontal bar chart titled ‘Accessibility Issues with DMH tracking services’, listing the following issues with their respective counts: ‘Confusing layout or navigation’ (11), ‘Inaccessible graphs or visual data representations’ (11), ‘Lack of screen reader support’ (9), ‘Timing out during data entry’ (6), ‘Poor color contrast’ (4), ‘Overly complicated security features’ (2), and ‘Other’ (2).}
        \label{subfig:access-issues}
    \end{subfigure}
    \hfill
    \begin{subfigure}[b]{0.8\textwidth}
        \centering
        \includegraphics[width=\textwidth]{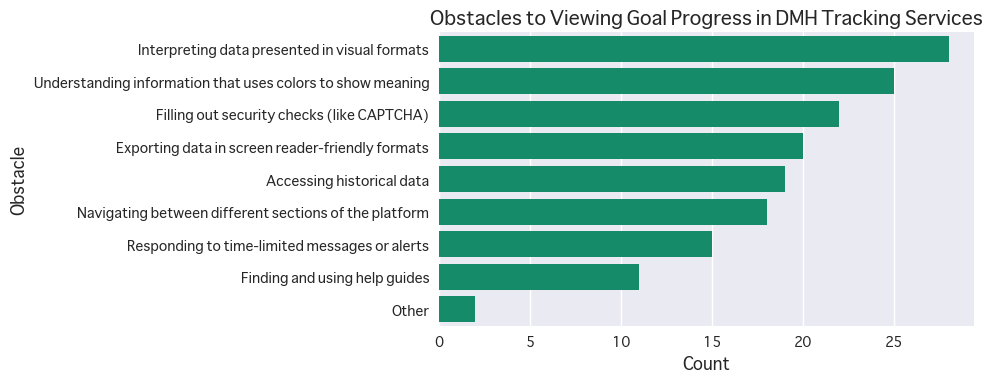}
        \caption{Reported obstacles to viewing goal progress with DMH tracking services.}
        \Description{Horizontal bar chart titled ‘Obstacles to Viewing Goal Progress in DMH Tracking Services’, listing the following obstacles with their respective counts: ‘Interpreting data presented in visual formats’ (25), ‘Understanding information that uses colors to show meaning’ (22), ‘Filling out security checks (like CAPTCHA)’ (18), ‘Exporting data in screen reader-friendly formats’ (16), ‘Accessing historical data’ (15), ‘Navigating between different sections of the platform’ (14), ‘Responding to time-limited messages or alerts’ (10), ‘Finding and using help guides’ (9), and ‘Other’ (2).}
        \label{subfig:obstacles}
    \end{subfigure}
    \hfill
    \begin{subfigure}[b]{0.8\textwidth}
        \centering
        \includegraphics[width=\textwidth]{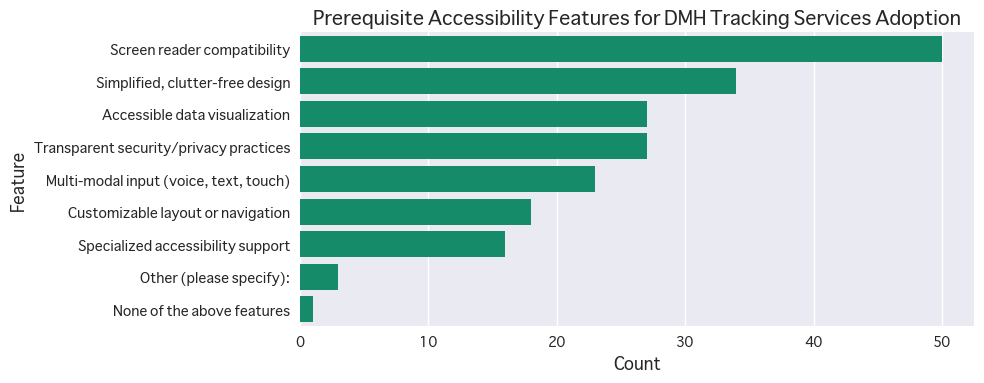}
        \caption{Reported prerequisite access features for DMH tracking service adoption.}
        \Description{Horizontal bar chart showing prerequisite accessibility features for DMH tracking services adoption. Y-axis lists features, X-axis shows count. Features in descending order: Screen reader compatibility (50), Simplified clutter-free design (40), Accessible data visualization (30), Transparent security/privacy practices (30), Multi-modal input (25), Customizable layout or navigation (20), Specialized accessibility support (15), Other (5), None of the above features (2).}
        \label{subfig:prereq-features}
    \end{subfigure}
    
    \caption{Participants' specific usability challenges and requirements for DMH tracking services.}
    \Description{Three horizontal bar charts showing participants' Reported obstacles to viewing goal progress(center), reported access challenges (top), and reported prerequisite accessibility features before DMH tracking service adoption (bottom).}
    \label{fig:dmh-specific-issues}
\end{figure}

%% file: references/a11y_framework.bib
@inproceedings{mankoffDisabilityStudiesSource2010,
  title = {Disability Studies as a Source of Critical Inquiry for the Field of Assistive Technology},
  booktitle = {Proceedings of the 12th International {{ACM SIGACCESS}} Conference on {{Computers}} and Accessibility},
  author = {Mankoff, Jennifer and Hayes, Gillian R. and Kasnitz, Devva},
  year = {2010},
  month = oct,
  series = {{{ASSETS}} '10},
  pages = {3--10},
  publisher = {Association for Computing Machinery},
  address = {New York, NY, USA},
  doi = {10.1145/1878803.1878807},
  urldate = {2022-11-06},
  isbn = {978-1-60558-881-0}
}

@inproceedings{sharifShouldSayDisabled2022,
  title = {Should {{I Say}} ``{{Disabled People}}'' or ``{{People}} with {{Disabilities}}''? {{Language Preferences}} of {{Disabled People Between Identity-}} and {{Person-First Language}}},
  shorttitle = {Should {{I Say}} ``{{Disabled People}}'' or ``{{People}} with {{Disabilities}}''?},
  booktitle = {Proceedings of the 24th {{International ACM SIGACCESS Conference}} on {{Computers}} and {{Accessibility}}},
  author = {Sharif, Ather and McCall, Aedan Liam and Bolante, Kianna Roces},
  year = {2022},
  month = oct,
  series = {{{ASSETS}} '22},
  pages = {1--18},
  publisher = {Association for Computing Machinery},
  address = {New York, NY, USA},
  doi = {10.1145/3517428.3544813},
  urldate = {2024-06-23},
  isbn = {978-1-4503-9258-7}
}

@inproceedings{spielNothingUsUs2020,
  title = {Nothing {{About Us Without Us}}: {{Investigating}} the {{Role}} of {{Critical Disability Studies}} in {{HCI}}},
  shorttitle = {Nothing {{About Us Without Us}}},
  booktitle = {Extended {{Abstracts}} of the 2020 {{CHI Conference}} on {{Human Factors}} in {{Computing Systems}}},
  author = {Spiel, Katta and Gerling, Kathrin and Bennett, Cynthia L. and Brul{\'e}, Emeline and Williams, Rua M. and Rode, Jennifer and Mankoff, Jennifer},
  year = {2020},
  month = apr,
  series = {{{CHI EA}} '20},
  pages = {1--8},
  publisher = {Association for Computing Machinery},
  address = {New York, NY, USA},
  doi = {10.1145/3334480.3375150},
  urldate = {2024-07-22},
  isbn = {978-1-4503-6819-3}
}


%% file: references/references.bib
@incollection{cooper_thematic_2012,
	address = {Washington},
	title = {Thematic analysis.},
	isbn = {978-1-4338-1005-3},
	url = {https://content.apa.org/books/13620-004},
	doi = {10.1037/13620-004},
	language = {en},
	urldate = {2026-07-18},
	booktitle = {{APA} handbook of research methods in psychology, {Vol} 2: {Research} designs: {Quantitative}, qualitative, neuropsychological, and biological.},
	publisher = {American Psychological Association},
	author = {Braun, Virginia and Clarke, Victoria},
	editor = {Cooper, Harris and Camic, Paul M. and Long, Debra L. and Panter, A. T. and Rindskopf, David and Sher, Kenneth J.},
	year = {2012},
	doi = {10.1037/13620-004},
	pages = {57--71},
}

@article{zhang_designing_2024,
	title = {Designing {Accessible} {Content} {Creation} {Support} with {Blind} and {Low} {Vision} {Creators}},
	issn = {1558-2337},
	url = {https://doi.org/10.1145/3654768.3654775},
	doi = {10.1145/3654768.3654775},
	number = {137},
	urldate = {2026-07-18},
	journal = {ACM SIGACCESS Accessibility and Computing},
	author = {Zhang, Lotus},
	month = mar,
	year = {2024},
	pages = {7:1},
}

@article{mcdonnall_availability_2017,
	title = {Availability of {Mental} {Health} {Services} for {Individuals} {Who} {Are} {Deaf} or {Deaf}-{Blind}},
	volume = {16},
	issn = {1536-7118},
	doi = {10.1080/1536710X.2017.1260515},
	language = {eng},
	number = {1},
	journal = {Journal of Social Work in Disability \& Rehabilitation},
	author = {McDonnall, Michele C. and Crudden, Adele and LeJeune, B. J. and Steverson, Anne Carter},
	year = {2017},
	pages = {1--13},
}

@article{sheikhWearableEnvironmentalSmartphoneBased2021,
	title = {Wearable, {Environmental}, and {Smartphone}-{Based} {Passive} {Sensing} for {Mental} {Health} {Monitoring}},
	volume = {3},
	issn = {2673-253X},
	url = {https://www.frontiersin.org/journals/digital-health/articles/10.3389/fdgth.2021.662811/full},
	doi = {10.3389/fdgth.2021.662811},
	language = {English},
	urldate = {2024-07-07},
	journal = {Frontiers in Digital Health},
	publisher = {Frontiers},
	author = {Sheikh, Mahsa and Qassem, M. and Kyriacou, Panicos A.},
	month = apr,
	year = {2021},
	pages = {21},
}

@inproceedings{radford_robust_2023,
	address = {Honolulu, Hawaii},
	title = {Robust {Speech} {Recognition} via {Large}-{Scale} {Weak} {Supervision}},
	issn = {2640-3498},
	url = {https://proceedings.mlr.press/v202/radford23a.html},
	language = {en},
	urldate = {2026-07-17},
	booktitle = {Proceedings of the 40th {International} {Conference} on {Machine} {Learning}},
	publisher = {PMLR},
	author = {Radford, Alec and Kim, Jong Wook and Xu, Tao and Brockman, Greg and Mcleavey, Christine and Sutskever, Ilya},
	month = jul,
	year = {2023},
	pages = {28492--28518},
}

@incollection{mouloudj_factors_2023,
	address = {Hershey, Pennsylvania},
	title = {Factors {Influencing} the {Adoption} of {Digital} {Health} {Apps}: {An} {Extended} {Technology} {Acceptance} {Model} ({TAM})},
	copyright = {Access limited to members},
	isbn = {978-1-6684-8337-4},
	shorttitle = {Factors {Influencing} the {Adoption} of {Digital} {Health} {Apps}},
	url = {https://www.igi-global.com/chapter/factors-influencing-the-adoption-of-digital-health-apps/www.igi-global.com/chapter/factors-influencing-the-adoption-of-digital-health-apps/323782},
	doi = {10.4018/978-1-6684-8337-4.ch007},
	language = {en},
	urldate = {2026-07-17},
	booktitle = {Integrating {Digital} {Health} {Strategies} for {Effective} {Administration}},
	publisher = {IGI Global Scientific Publishing},
	author = {Mouloudj, Kamel and Bouarar, Ahmed Chemseddine and Asanza, Dachel Martínez and Saadaoui, Linda and Mouloudj, Smail and Njoku, Anuli U. and Evans, Marian A. and Bouarar, Achouak},
	year = {2023},
	doi = {10.4018/978-1-6684-8337-4.ch007},
	pages = {116--132},
}

@book{creswell_designing_2017,
	address = {Thousand Oaks, CA},
	title = {Designing and {Conducting} {Mixed} {Methods} {Research}},
	isbn = {978-1-4833-4701-1},
	language = {en},
	publisher = {SAGE Publications},
	author = {Creswell, John W. and Clark, Vicki L. Plano},
	month = aug,
	year = {2017},
	note = {Google-Books-ID: BXEzDwAAQBAJ},
}

@article{norman_ehealth_2006,
	title = {{eHealth} {Literacy}: {Essential} {Skills} for {Consumer} {Health} in a {Networked} {World}},
	volume = {8},
	shorttitle = {{eHealth} {Literacy}},
	url = {https://www.jmir.org/2006/2/e9},
	doi = {10.2196/jmir.8.2.e9},
	language = {EN},
	number = {2},
	urldate = {2026-07-17},
	journal = {Journal of Medical Internet Research},
	publisher = {JMIR Publications Inc., Toronto, Canada},
	author = {Norman, Cameron D. and Skinner, Harvey A.},
	month = jun,
	year = {2006},
	pages = {e506},
}

@inproceedings{soubutts_challenges_2024,
	address = {New York, NY, USA},
	series = {{CHI} '24},
	title = {Challenges and {Opportunities} for the {Design} of {Inclusive} {Digital} {Mental} {Health} {Tools}: {Understanding} {Culturally} {Diverse} {Young} {People}'s {Experiences}},
	isbn = {979-8-4007-0330-0},
	shorttitle = {Challenges and {Opportunities} for the {Design} of {Inclusive} {Digital} {Mental} {Health} {Tools}},
	url = {https://dl.acm.org/doi/10.1145/3613904.3642641},
	doi = {10.1145/3613904.3642641},
	urldate = {2026-07-17},
	booktitle = {Proceedings of the 2024 {CHI} {Conference} on {Human} {Factors} in {Computing} {Systems}},
	publisher = {Association for Computing Machinery},
	author = {Soubutts, Ewan and Shrestha, Pranita and Davidson, Brittany I and Qu, Chengcheng and Mindel, Charlotte and Sefi, Aaron and Marshall, Paul and Mcnaney, Roisin},
	month = may,
	year = {2024},
	pages = {1--16},
}

@article{chaudary_teleguidance-based_2021,
	title = {Teleguidance-based remote navigation assistance for visually impaired and blind people—usability and user experience},
	volume = {27},
	issn = {1359-4338},
	url = {https://doi.org/10.1007/s10055-021-00536-z},
	doi = {10.1007/s10055-021-00536-z},
	number = {1},
	urldate = {2026-07-17},
	journal = {Virtual Reality},
	author = {Chaudary, Babar and Pohjolainen, Sami and Aziz, Saima and Arhippainen, Leena and Pulli, Petri},
	month = may,
	year = {2021},
	pages = {141--158},
}

@inproceedings{smith_asset_2018,
	address = {New York, NY, USA},
	series = {{SIGCSE} '18},
	title = {Asset {Maps}: {A} {Simple} {Tool} for {Recruiting} and {Retaining} {Underrepresented} {Populations} in {Computer} {Science} ({Abstract} {Only})},
	isbn = {978-1-4503-5103-4},
	shorttitle = {Asset {Maps}},
	url = {https://doi.org/10.1145/3159450.3162230},
	doi = {10.1145/3159450.3162230},
	urldate = {2026-07-17},
	booktitle = {Proceedings of the 49th {ACM} {Technical} {Symposium} on {Computer} {Science} {Education}},
	publisher = {Association for Computing Machinery},
	author = {Smith, Adrienne and Zulli, Rebecca},
	month = feb,
	year = {2018},
	pages = {1109},
}

@article{progga_womens_2025,
	title = {Women's {Perspectives} and {Challenges} in {Adopting} {Perinatal} {Mental} {Health} {Technologies}},
	volume = {9},
	url = {https://dl.acm.org/doi/10.1145/3701217},
	doi = {10.1145/3701217},
	number = {1},
	urldate = {2026-07-17},
	journal = {Proceedings of the ACM on Human-Computer Interaction},
	author = {Progga, Farhat Tasnim and Rubya, Sabirat},
	month = jan,
	year = {2025},
	pages = {GROUP38:1--GROUP38:30},
}

@article{mair_understanding_2025,
	title = {Understanding and overcoming barriers to digital health adoption: a patient and public involvement study},
	volume = {15},
	issn = {1869-6716},
	shorttitle = {Understanding and overcoming barriers to digital health adoption},
	url = {https://doi.org/10.1093/tbm/ibaf010},
	doi = {10.1093/tbm/ibaf010},
	number = {1},
	urldate = {2026-07-17},
	journal = {Translational Behavioral Medicine},
	author = {Mair, Jacqueline Louise and Hashim, Jumana and Thai, Linh and Tai, E Shyong and Ryan, Jillian C and Kowatsch, Tobias and Müller-Riemenschneider, Falk and Edney, Sarah Martine},
	month = jan,
	year = {2025},
	pages = {ibaf010},
}

@article{kruskal_use_1952,
	title = {Use of {Ranks} in {One}-{Criterion} {Variance} {Analysis}},
	volume = {47},
	issn = {0162-1459},
	url = {https://doi.org/10.1080/01621459.1952.10483441},
	doi = {10.1080/01621459.1952.10483441},
	number = {260},
	urldate = {2026-07-17},
	journal = {Journal of the American Statistical Association},
	publisher = {Taylor \& Francis},
	author = {Kruskal, William H. and Wallis, W. Allen},
	month = dec,
	year = {1952},
	note = {\_eprint: https://doi.org/10.1080/01621459.1952.10483441},
	pages = {583--621},
}

@inproceedings{kevin_chapman_mental_2019,
	address = {New York, NY, USA},
	series = {{SIGUCCS} '19},
	title = {Mental {Health} in the {IT} {Workplace}},
	isbn = {978-1-4503-5774-6},
	url = {https://dl.acm.org/doi/10.1145/3347709.3347828},
	doi = {10.1145/3347709.3347828},
	urldate = {2026-07-17},
	booktitle = {Proceedings of the 2019 {ACM} {SIGUCCS} {Annual} {Conference}},
	publisher = {Association for Computing Machinery},
	author = {Kevin Chapman, R},
	month = oct,
	year = {2019},
	pages = {209},
}

@article{braun_using_2006,
	title = {Using thematic analysis in psychology},
	volume = {3},
	issn = {1478-0887},
	url = {https://doi.org/10.1191/1478088706qp063oa},
	doi = {10.1191/1478088706qp063oa},
	number = {2},
	urldate = {2026-07-17},
	journal = {Qualitative Research in Psychology},
	publisher = {Routledge},
	author = {Braun, Virginia and Clarke, Victoria},
	month = jan,
	year = {2006},
	note = {\_eprint: https://doi.org/10.1191/1478088706qp063oa},
	pages = {77--101},
}

@article{davis_perceived_1989,
	title = {Perceived {Usefulness}, {Perceived} {Ease} of {Use}, and {User} {Acceptance} of {Information} {Technology}},
	volume = {13},
	issn = {0276-7783},
	url = {https://dx.doi.org/10.2307/249008},
	doi = {10.2307/249008},
	language = {en},
	number = {3},
	urldate = {2026-07-17},
	journal = {Management Information Systems Quarterly},
	publisher = {MIS Quarterly},
	author = {Davis, Fred D.},
	month = sep,
	year = {1989},
	pages = {319--340},
}

@article{lee_personal_2023,
	title = {Personal {Health} {Data} {Tracking} by {Blind} and {Low}-{Vision} {People}: {Survey} {Study}},
	volume = {25},
	issn = {1438-8871},
	shorttitle = {Personal {Health} {Data} {Tracking} by {Blind} and {Low}-{Vision} {People}},
	doi = {10.2196/43917},
	language = {eng},
	journal = {Journal of Medical Internet Research},
	author = {Lee, Jarrett G. W. and Lee, Kyungyeon and Lee, Bongshin and Choi, Soyoung and Seo, JooYoung and Choe, Eun Kyoung},
	month = may,
	year = {2023},
	pages = {e43917},
}

@article{choi_exploring_2024,
	title = {Exploring {mHealth} design opportunities for blind and visually impaired older users},
	volume = {10},
	issn = {2306-9740},
	doi = {10.21037/mhealth-23-65},
	language = {eng},
	journal = {mHealth},
	author = {Choi, Soyoung and Chlebek, Christian Joseph},
	year = {2024},
	pages = {17},
}

@inproceedings{khan_sighted_2025,
	address = {New York, NY, USA},
	series = {{CHI} {EA} '25},
	title = {"{Sighted} {People} {Have} {Their} {Pick} {Of} {The} {Litter}": {Unpacking} {The} {Need} {For} {Digital} {Mental} {Health} ({DMH}) {Tracking} {Services} {With} {And} {For} {The} {Blind} {Community}},
	isbn = {979-8-4007-1395-8},
	shorttitle = {"{Sighted} {People} {Have} {Their} {Pick} {Of} {The} {Litter}"},
	url = {https://dl.acm.org/doi/10.1145/3706599.3719817},
	doi = {10.1145/3706599.3719817},
	urldate = {2026-07-17},
	booktitle = {Proceedings of the {Extended} {Abstracts} of the {CHI} {Conference} on {Human} {Factors} in {Computing} {Systems}},
	publisher = {Association for Computing Machinery},
	author = {Khan, Omar and Seo, JooYoung},
	month = apr,
	year = {2025},
	pages = {1--13},
}

@article{nutbeamHealthLiteracyPublic2000,
	title = {Health literacy as a public health goal: a challenge for contemporary health education and communication strategies into the 21st century},
	volume = {15},
	issn = {0957-4824},
	shorttitle = {Health literacy as a public health goal},
	url = {https://doi.org/10.1093/heapro/15.3.259},
	doi = {10.1093/heapro/15.3.259},
	number = {3},
	urldate = {2026-04-23},
	journal = {Health Promotion International},
	author = {Nutbeam, Don},
	month = sep,
	year = {2000},
	pages = {259--267},
}

@article{fornacon-woodUnderstandingDifferencesBayesian2022,
	title = {Understanding the {Differences} {Between} {Bayesian} and {Frequentist} {Statistics}},
	volume = {112},
	issn = {0360-3016},
	url = {https://www.redjournal.org/article/S0360-3016(21)03256-9/fulltext},
	doi = {10.1016/j.ijrobp.2021.12.011},
	language = {English},
	number = {5},
	urldate = {2024-12-24},
	journal = {International Journal of Radiation Oncology, Biology, Physics},
	publisher = {Elsevier},
	author = {Fornacon-Wood, Isabella and Mistry, Hitesh and Johnson-Hart, Corinne and Faivre-Finn, Corinne and O'Connor, James P. B. and Price, Gareth J.},
	month = apr,
	year = {2022},
	pages = {1076--1082},
}

@article{crossDigitalCumulativeComplexity2024,
	title = {The digital cumulative complexity model: a framework for improving engagement in digital mental health interventions},
	volume = {15},
	issn = {1664-0640},
	shorttitle = {The digital cumulative complexity model},
	url = {https://www.frontiersin.org/journals/psychiatry/articles/10.3389/fpsyt.2024.1382726/full},
	doi = {10.3389/fpsyt.2024.1382726},
	language = {English},
	urldate = {2024-12-24},
	journal = {Frontiers in Psychiatry},
	publisher = {Frontiers},
	author = {Cross, Shane P. and Alvarez-Jimenez, Mario},
	month = sep,
	year = {2024},
}

@article{chenHybridCareMental2024,
	title = {Hybrid care in mental health: a framework for understanding care, research, and future opportunities},
	volume = {2},
	copyright = {2024 The Author(s)},
	issn = {2948-1570},
	shorttitle = {Hybrid care in mental health},
	url = {https://www.nature.com/articles/s44277-024-00016-7},
	doi = {10.1038/s44277-024-00016-7},
	language = {en},
	number = {1},
	urldate = {2024-12-24},
	journal = {NPP—Digital Psychiatry and Neuroscience},
	publisher = {Nature Publishing Group},
	author = {Chen, Kelly and Huang, Jack J. and Torous, John},
	month = oct,
	year = {2024},
	pages = {1--4},
}

@article{pinedaUpdatedTaxonomyDigital2023,
	title = {Updated taxonomy of digital mental health interventions: a conceptual framework},
	volume = {9},
	issn = {2306-9740},
	shorttitle = {Updated taxonomy of digital mental health interventions},
	url = {https://www.ncbi.nlm.nih.gov/pmc/articles/PMC10364001/},
	doi = {10.21037/mhealth-23-6},
	urldate = {2024-12-24},
	journal = {mHealth},
	author = {Pineda, Blanca S. and Mejia, Rosalva and Qin, Yuanzhi and Martinez, Julian and Delgadillo, Lizbet G. and Muñoz, Ricardo F.},
	month = jun,
	year = {2023},
	pages = {28},
}

@article{norgaardEhealthLiteracyFramework2015,
	title = {The e-health literacy framework: {A} conceptual framework for characterizing e-health users and their interaction with e-health systems},
	volume = {7},
	shorttitle = {The e-health literacy framework},
	journal = {Knowledge Management and E-Learning},
	author = {Norgaard, Ole and Furstrand, Dorthe and Klokker, Louise and Karnoe Knudsen, Astrid and Batterham, Roy and Kayser, Lars and Osborne, Richard},
	month = dec,
	year = {2015},
	pages = {522--40},
}

@inproceedings{alslaityInsightsLongitudinalEvaluation2022,
	address = {New York, NY, USA},
	series = {{CHI} {EA} '22},
	title = {Insights {From} {Longitudinal} {Evaluation} of {Moodie} {Mental} {Health} {App}},
	isbn = {978-1-4503-9156-6},
	url = {https://doi.org/10.1145/3491101.3519851},
	doi = {10.1145/3491101.3519851},
	urldate = {2024-07-10},
	booktitle = {Extended {Abstracts} of the 2022 {CHI} {Conference} on {Human} {Factors} in {Computing} {Systems}},
	publisher = {Association for Computing Machinery},
	author = {Alslaity, Alaa and Chan, Gerry and Orji, Rita and Wilson, Richard},
	month = apr,
	year = {2022},
	pages = {1--8},
}

@inproceedings{barryMHealthMaternalMental2017,
	address = {New York, NY, USA},
	series = {{CHI} '17},
	title = {{mHealth} for {Maternal} {Mental} {Health}: {Everyday} {Wisdom} in {Ethical} {Design}},
	isbn = {978-1-4503-4655-9},
	shorttitle = {{mHealth} for {Maternal} {Mental} {Health}},
	url = {https://doi.org/10.1145/3025453.3025918},
	doi = {10.1145/3025453.3025918},
	urldate = {2024-07-10},
	booktitle = {Proceedings of the 2017 {CHI} {Conference} on {Human} {Factors} in {Computing} {Systems}},
	publisher = {Association for Computing Machinery},
	author = {Barry, Marguerite and Doherty, Kevin and Marcano Belisario, Jose and Car, Josip and Morrison, Cecily and Doherty, Gavin},
	month = may,
	year = {2017},
	pages = {2708--2756},
}

@inproceedings{ausmanArtificialIntelligencesImpact2019,
	address = {New York, NY, USA},
	series = {{AIES} '19},
	title = {Artificial {Intelligence}'s {Impact} on {Mental} {Health} {Treatments}},
	isbn = {978-1-4503-6324-2},
	url = {https://doi.org/10.1145/3306618.3314311},
	doi = {10.1145/3306618.3314311},
	urldate = {2024-09-09},
	booktitle = {Proceedings of the 2019 {AAAI}/{ACM} {Conference} on {AI}, {Ethics}, and {Society}},
	publisher = {Association for Computing Machinery},
	author = {Ausman, Michelle C.},
	month = jan,
	year = {2019},
	pages = {533--534},
}

@article{attkissonClientSatisfactionQuestionnaire1982,
	title = {The client satisfaction questionnaire. {Psychometric} properties and correlations with service utilization and psychotherapy outcome},
	volume = {5},
	issn = {0149-7189},
	doi = {10.1016/0149-7189(82)90074-x},
	language = {eng},
	number = {3},
	journal = {Evaluation and Program Planning},
	author = {Attkisson, C. C. and Zwick, R.},
	year = {1982},
	pages = {233--237},
}

@article{parmantoDevelopmentTelehealthUsability2016,
	title = {Development of the {Telehealth} {Usability} {Questionnaire} ({TUQ})},
	volume = {8},
	issn = {1945-2020},
	url = {https://www.ncbi.nlm.nih.gov/pmc/articles/PMC4985278/},
	doi = {10.5195/ijt.2016.6196},
	number = {1},
	urldate = {2024-08-26},
	journal = {International Journal of Telerehabilitation},
	author = {PARMANTO, BAMBANG and LEWIS, ALLEN NELSON and GRAHAM, KRISTIN M. and BERTOLET, MARNIE H.},
	month = jul,
	year = {2016},
	pages = {3--10},
}

@inproceedings{oewelApproachesTailoringBetweensession2024,
	address = {New York, NY, USA},
	series = {{CHI} '24},
	title = {Approaches for tailoring between-session mental health therapy activities},
	isbn = {979-8-4007-0330-0},
	url = {https://dl.acm.org/doi/10.1145/3613904.3642856},
	doi = {10.1145/3613904.3642856},
	urldate = {2024-08-22},
	booktitle = {Proceedings of the {CHI} {Conference} on {Human} {Factors} in {Computing} {Systems}},
	publisher = {Association for Computing Machinery},
	author = {Oewel, Bruna and Arean, Patricia Anne and Agapie, Elena},
	month = may,
	year = {2024},
	pages = {1--19},
}

@article{ayobiDigitalMentalHealth2022,
	title = {Digital {Mental} {Health} and {Social} {Connectedness}: {Experiences} of {Women} from {Refugee} {Backgrounds}},
	volume = {6},
	shorttitle = {Digital {Mental} {Health} and {Social} {Connectedness}},
	url = {https://dl.acm.org/doi/10.1145/3555620},
	doi = {10.1145/3555620},
	number = {CSCW2},
	urldate = {2024-08-16},
	journal = {Proc. ACM Hum.-Comput. Interact.},
	author = {Ayobi, Amid and Eardley, Rachel and Soubutts, Ewan and Gooberman-Hill, Rachael and Craddock, Ian and O'Kane, Aisling Ann},
	month = nov,
	year = {2022},
	pages = {507:1--507:27},
}

@inproceedings{coyleInteractionDesignEmotional2012,
	address = {New York, NY, USA},
	series = {{CHI} {EA} '12},
	title = {Interaction design and emotional wellbeing},
	isbn = {978-1-4503-1016-1},
	url = {https://dl.acm.org/doi/10.1145/2212776.2212718},
	doi = {10.1145/2212776.2212718},
	urldate = {2024-08-13},
	booktitle = {{CHI} '12 {Extended} {Abstracts} on {Human} {Factors} in {Computing} {Systems}},
	publisher = {Association for Computing Machinery},
	author = {Coyle, David and Linehan, Conor and Tang, Karen and Lindley, Sian},
	month = may,
	year = {2012},
	pages = {2775--2778},
}

@inproceedings{baghaeiTimeGetPersonal2020,
	address = {New York, NY, USA},
	series = {{CHI} {EA} '20},
	title = {Time to {Get} {Personal}: {Individualised} {Virtual} {Reality} for {Mental} {Health}},
	isbn = {978-1-4503-6819-3},
	shorttitle = {Time to {Get} {Personal}},
	url = {https://doi.org/10.1145/3334480.3382932},
	doi = {10.1145/3334480.3382932},
	urldate = {2024-08-12},
	booktitle = {Extended {Abstracts} of the 2020 {CHI} {Conference} on {Human} {Factors} in {Computing} {Systems}},
	publisher = {Association for Computing Machinery},
	author = {Baghaei, Nilufar and Stemmet, Lehan and Hlasnik, Andrej and Emanov, Konstantin and Hach, Sylvia and Naslund, John A. and Billinghurst, Mark and Khaliq, Imran and Liang, Hai-Ning},
	month = apr,
	year = {2020},
	pages = {1--9},
}

@inproceedings{feinbergZenVRDesignEvaluation2022,
	address = {New York, NY, USA},
	series = {{CHI} '22},
	title = {{ZenVR}: {Design} {Evaluation} of a {Virtual} {Reality} {Learning} {System} for {Meditation}},
	isbn = {978-1-4503-9157-3},
	shorttitle = {{ZenVR}},
	url = {https://dl.acm.org/doi/10.1145/3491102.3502035},
	doi = {10.1145/3491102.3502035},
	urldate = {2024-08-12},
	booktitle = {Proceedings of the 2022 {CHI} {Conference} on {Human} {Factors} in {Computing} {Systems}},
	publisher = {Association for Computing Machinery},
	author = {Feinberg, Rachel R. and Lakshmi, Udaya and Golino, Matthew J. and Arriaga, Rosa I.},
	month = apr,
	year = {2022},
	pages = {1--15},
}

@inproceedings{zhangDesigningEmotionalWellbeing2021,
	address = {New York, NY, USA},
	series = {{CHI} '21},
	title = {Designing for {Emotional} {Well}-being: {Integrating} {Persuasion} and {Customization} into {Mental} {Health} {Technologies}},
	isbn = {978-1-4503-8096-6},
	shorttitle = {Designing for {Emotional} {Well}-being},
	url = {https://doi.org/10.1145/3411764.3445771},
	doi = {10.1145/3411764.3445771},
	urldate = {2024-07-23},
	booktitle = {Proceedings of the 2021 {CHI} {Conference} on {Human} {Factors} in {Computing} {Systems}},
	publisher = {Association for Computing Machinery},
	author = {Zhang, Renwen and E. Ringland, Kathryn and Paan, Melina and C. Mohr, David and Reddy, Madhu},
	month = may,
	year = {2021},
	pages = {1--13},
}

@article{adamsAvailabilityAccessibilityMental2024,
	title = {Availability and {Accessibility} of {Mental} {Health} {Services} for {Youth}: {A} {Descriptive} {Survey} of {Safety}-{Net} {Health} {Centers} {During} the {COVID}-19 {Pandemic}},
	volume = {60},
	issn = {1573-2789},
	shorttitle = {Availability and {Accessibility} of {Mental} {Health} {Services} for {Youth}},
	url = {https://doi.org/10.1007/s10597-023-01127-9},
	doi = {10.1007/s10597-023-01127-9},
	language = {en},
	number = {1},
	urldate = {2024-06-23},
	journal = {Community Mental Health Journal},
	author = {Adams, Danielle R.},
	month = jan,
	year = {2024},
	pages = {88--97},
}

@inproceedings{bowie-dabreoUserPerspectivesEthical2022,
	address = {New York, NY, USA},
	series = {{CHI} '22},
	title = {User {Perspectives} and {Ethical} {Experiences} of {Apps} for {Depression}: {A} {Qualitative} {Analysis} of {User} {Reviews}},
	isbn = {978-1-4503-9157-3},
	shorttitle = {User {Perspectives} and {Ethical} {Experiences} of {Apps} for {Depression}},
	url = {https://doi.org/10.1145/3491102.3517498},
	doi = {10.1145/3491102.3517498},
	urldate = {2024-07-10},
	booktitle = {Proceedings of the 2022 {CHI} {Conference} on {Human} {Factors} in {Computing} {Systems}},
	publisher = {Association for Computing Machinery},
	author = {Bowie-DaBreo, Dionne and Sas, Corina and Iles-Smith, Heather and Sünram-Lea, Sandra},
	month = apr,
	year = {2022},
	pages = {1--24},
}

@inproceedings{cabreraEthicalDilemmasMental2023,
	address = {Cham},
	title = {Ethical {Dilemmas}, {Mental} {Health}, {Artificial} {Intelligence}, and {LLM}-{Based} {Chatbots}},
	isbn = {978-3-031-34960-7},
	doi = {10.1007/978-3-031-34960-7_22},
	language = {en},
	booktitle = {Bioinformatics and {Biomedical} {Engineering}},
	publisher = {Springer Nature Switzerland},
	author = {Cabrera, Johana and Loyola, M. Soledad and Magaña, Irene and Rojas, Rodrigo},
	editor = {Rojas, Ignacio and Valenzuela, Olga and Rojas Ruiz, Fernando and Herrera, Luis Javier and Ortuño, Francisco},
	year = {2023},
	pages = {313--326},
}

@inproceedings{calvoComputingMentalHealth2016,
	address = {New York, NY, USA},
	series = {{CHI} {EA} '16},
	title = {Computing in {Mental} {Health}},
	isbn = {978-1-4503-4082-3},
	url = {https://doi.org/10.1145/2851581.2856463},
	doi = {10.1145/2851581.2856463},
	urldate = {2024-07-10},
	booktitle = {Proceedings of the 2016 {CHI} {Conference} {Extended} {Abstracts} on {Human} {Factors} in {Computing} {Systems}},
	publisher = {Association for Computing Machinery},
	author = {Calvo, Rafael A. and Dinakar, Karthik and Picard, Rosalind and Maes, Pattie},
	month = may,
	year = {2016},
	pages = {3438--3445},
}

@inproceedings{calvoDesignWellbeingTools2019,
	address = {New York, NY, USA},
	series = {{CHI} {EA} '19},
	title = {Design for {Wellbeing} - {Tools} for {Research}, {Practice} and {Ethics}},
	isbn = {978-1-4503-5971-9},
	url = {https://doi.org/10.1145/3290607.3298800},
	doi = {10.1145/3290607.3298800},
	urldate = {2024-07-10},
	booktitle = {Extended {Abstracts} of the 2019 {CHI} {Conference} on {Human} {Factors} in {Computing} {Systems}},
	publisher = {Association for Computing Machinery},
	author = {Calvo, Rafael A. and Peters, Dorian},
	month = may,
	year = {2019},
	pages = {1--5},
}

@inproceedings{chapmanSociotechnicalConsiderationsAccessibility2024,
	address = {New York, NY, USA},
	series = {{WWW} '24},
	title = {Sociotechnical {Considerations} for {Accessibility} and {Equity} in {AI} for {Healthcare}},
	isbn = {979-8-4007-0172-6},
	url = {https://dl.acm.org/doi/10.1145/3589335.3651455},
	doi = {10.1145/3589335.3651455},
	urldate = {2024-06-23},
	booktitle = {Companion {Proceedings} of the {ACM} on {Web} {Conference} 2024},
	publisher = {Association for Computing Machinery},
	author = {Chapman, Adriane and Harrison, Chloe L and Jones, Caroline and Thornton, James and Worley, Rose and Wyatt, Jeremy C.},
	month = may,
	year = {2024},
	pages = {1158--1161},
}

@inproceedings{calvoPositiveComputingResearch2017,
	address = {New York, NY, USA},
	series = {{CHI} {EA} '17},
	title = {Positive {Computing}: {Research} \& {Practice} in {Wellbeing} {Technology}},
	isbn = {978-1-4503-4656-6},
	shorttitle = {Positive {Computing}},
	url = {https://doi.org/10.1145/3027063.3027099},
	doi = {10.1145/3027063.3027099},
	urldate = {2024-07-10},
	booktitle = {Proceedings of the 2017 {CHI} {Conference} {Extended} {Abstracts} on {Human} {Factors} in {Computing} {Systems}},
	publisher = {Association for Computing Machinery},
	author = {Calvo, Rafael A. and Peters, Dorian},
	month = may,
	year = {2017},
	pages = {1220--1223},
}

@inproceedings{daudenroquetEvaluatingMindfulnessMeditation2018,
	address = {New York, NY, USA},
	series = {{CHI} {EA} '18},
	title = {Evaluating {Mindfulness} {Meditation} {Apps}},
	isbn = {978-1-4503-5621-3},
	url = {https://doi.org/10.1145/3170427.3188616},
	doi = {10.1145/3170427.3188616},
	urldate = {2024-07-10},
	booktitle = {Extended {Abstracts} of the 2018 {CHI} {Conference} on {Human} {Factors} in {Computing} {Systems}},
	publisher = {Association for Computing Machinery},
	author = {Daudén Roquet, Claudia and Sas, Corina},
	month = apr,
	year = {2018},
	pages = {1--6},
}

@article{guMentalBlendEnhancingOnline2024,
	title = {{MentalBlend}: {Enhancing} {Online} {Mental} {Health} {Support} through the {Integration} of {LLMs} with {Psychological} {Counseling} {Theories}},
	volume = {46},
	shorttitle = {{MentalBlend}},
	url = {https://escholarship.org/uc/item/7dk883nx},
	language = {en},
	number = {0},
	urldate = {2024-06-24},
	journal = {Proceedings of the Annual Meeting of the Cognitive Science Society},
	author = {Gu, Ziyin and Zhu, Qingmeng},
	year = {2024},
}

@inproceedings{hassanUnveilingPrivacyMeasures2023,
	address = {New York, NY, USA},
	series = {{UbiComp}/{ISWC} '23 {Adjunct}},
	title = {Unveiling {Privacy} {Measures} in {Mental} {Health} {Applications}},
	isbn = {979-8-4007-0200-6},
	url = {https://doi.org/10.1145/3594739.3612879},
	doi = {10.1145/3594739.3612879},
	urldate = {2024-07-10},
	booktitle = {Adjunct {Proceedings} of the 2023 {ACM} {International} {Joint} {Conference} on {Pervasive} and {Ubiquitous} {Computing} \& the 2023 {ACM} {International} {Symposium} on {Wearable} {Computing}},
	publisher = {Association for Computing Machinery},
	author = {Hassan, Muhammad and Bashir, Masooda},
	month = oct,
	year = {2023},
	pages = {648--654},
}

@inproceedings{hoeferVisualizingUncertaintyMultiSource2022,
	address = {New York, NY, USA},
	series = {{CHI} {EA} '22},
	title = {Visualizing {Uncertainty} in {Multi}-{Source} {Mental} {Health} {Data}},
	isbn = {978-1-4503-9156-6},
	url = {https://doi.org/10.1145/3491101.3519844},
	doi = {10.1145/3491101.3519844},
	urldate = {2024-07-10},
	booktitle = {Extended {Abstracts} of the 2022 {CHI} {Conference} on {Human} {Factors} in {Computing} {Systems}},
	publisher = {Association for Computing Machinery},
	author = {Hoefer, Michael Jeffrey Daniel and Schumacher, Bryce E and Szafir, Danielle Albers and Voida, Stephen},
	month = apr,
	year = {2022},
	pages = {1--6},
}

@inproceedings{kangThisAppSaid2024,
	address = {New York, NY, USA},
	series = {{CHI} '24},
	title = {“{This} app said {I} had severe depression, and now {I} don’t know what to do”: the unintentional harms of mental health applications},
	isbn = {979-8-4007-0330-0},
	shorttitle = {“{This} app said {I} had severe depression, and now {I} don’t know what to do”},
	url = {https://dl.acm.org/doi/10.1145/3613904.3642178},
	doi = {10.1145/3613904.3642178},
	urldate = {2024-06-23},
	booktitle = {Proceedings of the {CHI} {Conference} on {Human} {Factors} in {Computing} {Systems}},
	publisher = {Association for Computing Machinery},
	author = {Kang, Rachael M. and Reynolds, Tera L.},
	month = may,
	year = {2024},
	pages = {1--17},
}

@inproceedings{kellyItsMissingMuch2021,
	address = {New York, NY, USA},
	series = {{CHI} '21},
	title = {“{It}’s {About} {Missing} {Much} {More} {Than} the {People}”: {How} {Students} use {Digital} {Technologies} to {Alleviate} {Homesickness}},
	isbn = {978-1-4503-8096-6},
	shorttitle = {“{It}’s {About} {Missing} {Much} {More} {Than} the {People}”},
	url = {https://doi.org/10.1145/3411764.3445362},
	doi = {10.1145/3411764.3445362},
	urldate = {2024-07-10},
	booktitle = {Proceedings of the 2021 {CHI} {Conference} on {Human} {Factors} in {Computing} {Systems}},
	publisher = {Association for Computing Machinery},
	author = {Kelly, Ryan M. and Cheng, Yueyang and McKay, Dana and Wadley, Greg and Buchanan, George},
	month = may,
	year = {2021},
	pages = {1--17},
}

@inproceedings{khooThatsKindSuspicious2024,
	address = {New York, NY, USA},
	series = {{CHI} '24},
	title = {“{That}’s {Kind} of {Sus}(picious)”: {The} {Comprehensiveness} of {Mental} {Health} {Application} {Users}’ {Privacy} and {Security} {Concerns}},
	isbn = {979-8-4007-0330-0},
	shorttitle = {“{That}’s {Kind} of {Sus}(picious)”},
	url = {https://doi.org/10.1145/3613904.3642705},
	doi = {10.1145/3613904.3642705},
	urldate = {2024-07-10},
	booktitle = {Proceedings of the {CHI} {Conference} on {Human} {Factors} in {Computing} {Systems}},
	publisher = {Association for Computing Machinery},
	author = {Khoo, Yi Xuan and Kang, Rachael M. and Reynolds, Tera L. and Mentis, Helena M.},
	month = may,
	year = {2024},
	pages = {1--16},
}

@article{leeIdentifyAdaptPersist2024,
	title = {Identify, {Adapt}, {Persist}: {The} {Journey} of {Blind} {Individuals} with {Personal} {Health} {Technologies}},
	volume = {8},
	shorttitle = {Identify, {Adapt}, {Persist}},
	url = {https://doi.org/10.1145/3659585},
	doi = {10.1145/3659585},
	number = {2},
	urldate = {2024-07-07},
	journal = {Proc. ACM Interact. Mob. Wearable Ubiquitous Technol.},
	author = {Lee, Jarrett G.W. and Lee, Bongshin and Choi, Soyoung and Seo, JooYoung and Choe, Eun Kyoung},
	month = may,
	year = {2024},
	pages = {51:1--51:21},
}

@inproceedings{kornfieldEnergyFiniteResource2020,
	address = {New York, NY, USA},
	series = {{CHI} '20},
	title = {"{Energy} is a {Finite} {Resource}": {Designing} {Technology} to {Support} {Individuals} across {Fluctuating} {Symptoms} of {Depression}},
	isbn = {978-1-4503-6708-0},
	shorttitle = {"{Energy} is a {Finite} {Resource}"},
	url = {https://doi.org/10.1145/3313831.3376309},
	doi = {10.1145/3313831.3376309},
	urldate = {2024-07-10},
	booktitle = {Proceedings of the 2020 {CHI} {Conference} on {Human} {Factors} in {Computing} {Systems}},
	publisher = {Association for Computing Machinery},
	author = {Kornfield, Rachel and Zhang, Renwen and Nicholas, Jennifer and Schueller, Stephen M. and Cambo, Scott A. and Mohr, David C. and Reddy, Madhu},
	month = apr,
	year = {2020},
	pages = {1--17},
}

@misc{liAutomaticEvaluationMental2024,
	title = {Automatic {Evaluation} for {Mental} {Health} {Counseling} using {LLMs}},
	url = {http://arxiv.org/abs/2402.11958},
	doi = {10.48550/arXiv.2402.11958},
	urldate = {2024-06-24},
	publisher = {arXiv},
	author = {Li, Anqi and Lu, Yu and Song, Nirui and Zhang, Shuai and Ma, Lizhi and Lan, Zhenzhong},
	month = feb,
	year = {2024},
	note = {arXiv:2402.11958 [cs]},
}

@inproceedings{markumDigitalTechnologyMeditative2020,
	address = {New York, NY, USA},
	series = {{CHI} '20},
	title = {Digital {Technology}, {Meditative} and {Contemplative} {Practices}, and {Transcendent} {Experiences}},
	isbn = {978-1-4503-6708-0},
	url = {https://doi.org/10.1145/3313831.3376356},
	doi = {10.1145/3313831.3376356},
	urldate = {2024-07-10},
	booktitle = {Proceedings of the 2020 {CHI} {Conference} on {Human} {Factors} in {Computing} {Systems}},
	publisher = {Association for Computing Machinery},
	author = {Markum, Robert B. and Toyama, Kentaro},
	month = apr,
	year = {2020},
	pages = {1--14},
}

@inproceedings{meinlschmidtMentalHealthMetaverse2023,
	address = {New York, NY, USA},
	series = {{CHI} {EA} '23},
	title = {Mental {Health} and the {Metaverse}: {Ample} {Opportunities} or {Alarming} {Threats} for {Mental} {Health} in {Immersive} {Worlds}?},
	isbn = {978-1-4503-9422-2},
	shorttitle = {Mental {Health} and the {Metaverse}},
	url = {https://doi.org/10.1145/3544549.3583750},
	doi = {10.1145/3544549.3583750},
	urldate = {2024-06-25},
	booktitle = {Extended {Abstracts} of the 2023 {CHI} {Conference} on {Human} {Factors} in {Computing} {Systems}},
	publisher = {Association for Computing Machinery},
	author = {Meinlschmidt, Gunther and Herta, Stefanie and Germann, Stefan and Chee Pui Khei, Cliona and Klöss, Sebastian and Borrmann, Moritz},
	month = apr,
	year = {2023},
	pages = {1--5},
}

@article{nepalCurrentPracticesMental2021,
	title = {Current practices in mental health sensing},
	volume = {28},
	issn = {1528-4972},
	url = {https://doi.org/10.1145/3481829},
	doi = {10.1145/3481829},
	number = {1},
	urldate = {2024-07-10},
	journal = {XRDS},
	author = {Nepal, Subigya and Wang, Weichen and Sharma, Bishal and Paudel, Prabesh},
	month = sep,
	year = {2021},
	pages = {28--33},
}

@inproceedings{olearySuddenlyWeGot2018,
	address = {New York, NY, USA},
	series = {{CHI} '18},
	title = {“{Suddenly}, we got to become therapists for each other”: {Designing} {Peer} {Support} {Chats} for {Mental} {Health}},
	isbn = {978-1-4503-5620-6},
	shorttitle = {“{Suddenly}, we got to become therapists for each other”},
	url = {https://doi.org/10.1145/3173574.3173905},
	doi = {10.1145/3173574.3173905},
	urldate = {2024-07-10},
	booktitle = {Proceedings of the 2018 {CHI} {Conference} on {Human} {Factors} in {Computing} {Systems}},
	publisher = {Association for Computing Machinery},
	author = {O'Leary, Kathleen and Schueller, Stephen M. and Wobbrock, Jacob O. and Pratt, Wanda},
	month = apr,
	year = {2018},
	pages = {1--14},
}

@inproceedings{oguamanamIntersectionalLookUse2023,
	address = {New York, NY, USA},
	series = {{CHI} '23},
	title = {An {Intersectional} {Look} at {Use} of and {Satisfaction} with {Digital} {Mental} {Health} {Platforms}: {A} {Survey} of {Perinatal} {Black} {Women}},
	isbn = {978-1-4503-9421-5},
	shorttitle = {An {Intersectional} {Look} at {Use} of and {Satisfaction} with {Digital} {Mental} {Health} {Platforms}},
	url = {https://doi.org/10.1145/3544548.3581475},
	doi = {10.1145/3544548.3581475},
	urldate = {2024-07-10},
	booktitle = {Proceedings of the 2023 {CHI} {Conference} on {Human} {Factors} in {Computing} {Systems}},
	publisher = {Association for Computing Machinery},
	author = {Oguamanam, Vanessa O. and Hernandez, Natalie and Chandler, Rasheeta and Guillaume, Dominique and Mckeever, Kai and Allen, Morgan and Mohammed, Sabreen and Parker, Andrea G},
	month = apr,
	year = {2023},
	pages = {1--20},
}

@article{pandeyMentalHealthEvaluation2023,
	title = {Mental {Health} {Evaluation} and {Assistance} for {Visually} {Impaired} {People}},
	volume = {10},
	copyright = {Copyright (c) 2023 Dr. Kavita Pandey, Dr. Dhiraj Pandey},
	issn = {2032-9407},
	url = {https://publications.eai.eu/index.php/sis/article/view/2931},
	doi = {10.4108/eetsis.vi.2931},
	language = {en},
	number = {4},
	urldate = {2024-07-07},
	journal = {EAI Endorsed Transactions on Scalable Information Systems},
	author = {Pandey, Kavita and Pandey, Dhiraj},
	month = apr,
	year = {2023},
	note = {Number: 4},
	pages = {e6--e6},
}

@inproceedings{rectorExploringOpportunitiesChallenges2015,
	address = {New York, NY, USA},
	series = {{ASSETS} '15},
	title = {Exploring the {Opportunities} and {Challenges} with {Exercise} {Technologies} for {People} who are {Blind} or {Low}-{Vision}},
	isbn = {978-1-4503-3400-6},
	url = {https://doi.org/10.1145/2700648.2809846},
	doi = {10.1145/2700648.2809846},
	urldate = {2024-07-07},
	booktitle = {Proceedings of the 17th {International} {ACM} {SIGACCESS} {Conference} on {Computers} \& {Accessibility}},
	publisher = {Association for Computing Machinery},
	author = {Rector, Kyle and Milne, Lauren and Ladner, Richard E. and Friedman, Batya and Kientz, Julie A.},
	month = oct,
	year = {2015},
	pages = {203--214},
}

@article{robledoyamamotoTherapyTeletherapyRelocating2021,
	title = {From {Therapy} to {Teletherapy}: {Relocating} {Mental} {Health} {Services} {Online}},
	volume = {5},
	shorttitle = {From {Therapy} to {Teletherapy}},
	url = {https://dl.acm.org/doi/10.1145/3479508},
	doi = {10.1145/3479508},
	number = {CSCW2},
	urldate = {2024-06-24},
	journal = {Proceedings of the ACM on Human-Computer Interaction},
	author = {Robledo Yamamoto, Fujiko and Voida, Amy and Voida, Stephen},
	month = oct,
	year = {2021},
	pages = {364:1--364:30},
}

@article{sienCodesigningMentalHealth2023,
	title = {Co-designing {Mental} {Health} {Technologies} with {International} {University} {Students} in {Canada}},
	volume = {7},
	url = {https://doi.org/10.1145/3610049},
	doi = {10.1145/3610049},
	number = {CSCW2},
	urldate = {2024-06-24},
	journal = {Proceedings of the ACM on Human-Computer Interaction},
	author = {Sien, Sang-Wha and Ahn, Jessica Y. and McGrenere, Joanna},
	month = oct,
	year = {2023},
	pages = {258:1--258:25},
}

@inproceedings{slovakHCIContributionsMental2024,
	address = {New York, NY, USA},
	series = {{CHI} '24},
	title = {{HCI} {Contributions} in {Mental} {Health}: {A} {Modular} {Framework} to {Guide} {Psychosocial} {Intervention} {Design}},
	isbn = {979-8-4007-0330-0},
	shorttitle = {{HCI} {Contributions} in {Mental} {Health}},
	url = {https://doi.org/10.1145/3613904.3642624},
	doi = {10.1145/3613904.3642624},
	urldate = {2024-07-10},
	booktitle = {Proceedings of the {CHI} {Conference} on {Human} {Factors} in {Computing} {Systems}},
	publisher = {Association for Computing Machinery},
	author = {Slovak, Petr and Munson, Sean A.},
	month = may,
	year = {2024},
	pages = {1--21},
}

@inproceedings{soler-dominguezARCADIAGamifiedMixed2024,
	address = {New York, NY, USA},
	series = {{CHI} '24},
	title = {{ARCADIA}: {A} {Gamified} {Mixed} {Reality} {System} for {Emotional} {Regulation} and {Self}-{Compassion}},
	isbn = {979-8-4007-0330-0},
	shorttitle = {{ARCADIA}},
	url = {https://doi.org/10.1145/3613904.3642123},
	doi = {10.1145/3613904.3642123},
	urldate = {2024-07-10},
	booktitle = {Proceedings of the {CHI} {Conference} on {Human} {Factors} in {Computing} {Systems}},
	publisher = {Association for Computing Machinery},
	author = {Soler-Dominguez, Jose Luis and Navas-Medrano, Samuel and Pons, Patricia},
	month = may,
	year = {2024},
	pages = {1--17},
}

@inproceedings{sienDesigningInclusivityAccessibility2023,
	address = {New York, NY, USA},
	series = {{CHI} {EA} '23},
	title = {Designing for {Inclusivity} and {Accessibility} of {Mental} {Health} {Technologies}},
	isbn = {978-1-4503-9422-2},
	url = {https://doi.org/10.1145/3544549.3577038},
	doi = {10.1145/3544549.3577038},
	urldate = {2024-06-23},
	booktitle = {Extended {Abstracts} of the 2023 {CHI} {Conference} on {Human} {Factors} in {Computing} {Systems}},
	publisher = {Association for Computing Machinery},
	author = {Sien, Sang-Wha},
	month = apr,
	year = {2023},
	pages = {1--4},
}

@article{balcombeEvaluationUseDigital2023,
	title = {Evaluation of the {Use} of {Digital} {Mental} {Health} {Platforms} and {Interventions}: {Scoping} {Review}},
	volume = {20},
	copyright = {http://creativecommons.org/licenses/by/3.0/},
	issn = {1660-4601},
	shorttitle = {Evaluation of the {Use} of {Digital} {Mental} {Health} {Platforms} and {Interventions}},
	url = {https://www.mdpi.com/1660-4601/20/1/362},
	doi = {10.3390/ijerph20010362},
	language = {en},
	number = {1},
	urldate = {2024-06-24},
	journal = {International Journal of Environmental Research and Public Health},
	publisher = {Multidisciplinary Digital Publishing Institute},
	author = {Balcombe, Luke and De Leo, Diego},
	month = jan,
	year = {2023},
	note = {Number: 1},
	pages = {362},
}

@article{bunyiAccessibilityDigitalMental2021,
	title = {Accessibility and {Digital} {Mental} {Health}: {Considerations} for {More} {Accessible} and {Equitable} {Mental} {Health} {Apps}},
	volume = {3},
	issn = {2673-253X},
	shorttitle = {Accessibility and {Digital} {Mental} {Health}},
	url = {https://www.ncbi.nlm.nih.gov/pmc/articles/PMC8521906/},
	doi = {10.3389/fdgth.2021.742196},
	urldate = {2024-06-24},
	journal = {Frontiers in Digital Health},
	author = {Bunyi, John and Ringland, Kathryn E. and Schueller, Stephen M.},
	month = sep,
	year = {2021},
	pages = {742196},
}

@article{kohdaMentalHealthStatus2019,
	title = {Mental {Health} {Status} and {Related} {Factors} {Among} {Visually} {Impaired} {Athletes}},
	volume = {11},
	issn = {1918-3003},
	url = {https://www.ncbi.nlm.nih.gov/pmc/articles/PMC6879020/},
	doi = {10.14740/jocmr3984},
	number = {11},
	urldate = {2024-07-07},
	journal = {Journal of Clinical Medicine Research},
	author = {Kohda, Yasuko and Monma, Takafumi and Yamane, Maki and Mitsui, Toshihito and Ando, Kayoko and Jesmin, Subrina and Takeda, Fumi},
	month = nov,
	year = {2019},
	pages = {729--739},
}

@article{lattieOverviewRecommendationsMore2022,
	title = {An overview of and recommendations for more accessible digital mental health services},
	volume = {1},
	copyright = {2022 Springer Nature America, Inc.},
	issn = {2731-0574},
	url = {https://www.nature.com/articles/s44159-021-00003-1},
	doi = {10.1038/s44159-021-00003-1},
	language = {en},
	number = {2},
	urldate = {2024-06-23},
	journal = {Nature Reviews Psychology},
	publisher = {Nature Publishing Group},
	author = {Lattie, Emily G. and Stiles-Shields, Colleen and Graham, Andrea K.},
	month = feb,
	year = {2022},
	pages = {87--100},
}

@article{millerSelfMonitoringPhysicalActivity2022,
	title = {Self-{Monitoring} {Physical} {Activity}, {Diet}, and {Weight} {Among} {Adults} {Who} {Are} {Legally} {Blind}: {Exploratory} {Investigation}},
	volume = {9},
	copyright = {Unless stated otherwise, all articles are open-access distributed under the terms of the Creative Commons Attribution License (http://creativecommons.org/licenses/by/2.0/), which permits unrestricted use, distribution, and reproduction in any medium, provided the original work ("first published in the Journal of Medical Internet Research...") is properly cited with original URL and bibliographic citation information. The complete bibliographic information, a link to the original publication on http://www.jmir.org/, as well as this copyright and license information must be included.},
	shorttitle = {Self-{Monitoring} {Physical} {Activity}, {Diet}, and {Weight} {Among} {Adults} {Who} {Are} {Legally} {Blind}},
	url = {https://rehab.jmir.org/2022/4/e42923},
	doi = {10.2196/42923},
	language = {EN},
	number = {4},
	urldate = {2024-07-07},
	journal = {JMIR Rehabilitation and Assistive Technologies},
	publisher = {JMIR Publications Inc., Toronto, Canada},
	author = {Miller, Kamilla and Jerome, Gerald J.},
	month = dec,
	year = {2022},
	note = {Company: JMIR Rehabilitation and Assistive Technologies
Distributor: JMIR Rehabilitation and Assistive Technologies
Institution: JMIR Rehabilitation and Assistive Technologies
Label: JMIR Rehabilitation and Assistive Technologies},
	pages = {e42923},
}

@article{richardsonUnderutilizationMentalHealth2024,
	title = {The {Underutilization} of {Mental} {Health} {Care} {Services} in the {Lives} of {People} with {Blindness} or {Visual} {Impairment}: {A} {Literature} {Review} on {Rehabilitation} {Factors} {Toward} {Provision}},
	volume = {18},
	issn = {1177-5467},
	shorttitle = {The {Underutilization} of {Mental} {Health} {Care} {Services} in the {Lives} of {People} with {Blindness} or {Visual} {Impairment}},
	url = {https://www.tandfonline.com/doi/abs/10.2147/OPTH.S442430},
	doi = {10.2147/OPTH.S442430},
	urldate = {2024-06-24},
	journal = {Clinical Ophthalmology},
	publisher = {Dove Medical Press},
	author = {Richardson, Clairissa G},
	month = dec,
	year = {2024},
	note = {\_eprint: https://www.tandfonline.com/doi/pdf/10.2147/OPTH.S442430},
	pages = {953--980},
}


%% file: references/software.bib
@manual{ATLASTI_2024,
  title        = {ATLAS.ti (Version 24)},
  author       = {{ATLAS.ti Scientific Software Development GmbH}},
  organization = {{Lumivero}},
  year         = {2024},
  address      = {Berlin, Germany},
  url          = {https://atlasti.com}
}
